\documentclass[smallextended]{svjour3}
\smartqed  
\usepackage{subfigure}
\usepackage{natbib}
\usepackage{verbatim}
\usepackage{graphicx}
\usepackage{amssymb}  
\usepackage{amsmath}

\begin{document}

\title{Principal Component Analysis to correct data systematics. Case study: K2 light curves.}

\author{A. Petralia \and G. Micela} 
\institute{A. Petralia \at INAF-Osservatorio Astronomico di Palermo, Piazza del Parlamento 1, 90134 Palermo \\
              \email{antonino.petralia@inaf.it}             
           \and
           G.Micela \at
INAF-Osservatorio Astronomico di Palermo, Piazza del Parlamento 1, 90134 Palermo}

\date{Rec;Acc}

\maketitle

\begin{abstract}

 Instrumental data are affected by systematic effects that dominate the errors and can be relevant when searching for small signals. This is the case of {the} K2 mission, a follow up of the Kepler mission, that, after a failure on two reaction wheels, has lost its stability properties rising strongly the systematics in the light curves and reducing its photometric precision.  
 In this work, we have developed a general method to remove time related systematics from a set of light curves, that has been applied to K2 data.
 The method uses the Principal Component Analysis to retrieve the correlation between the light curves due to the systematics and to remove its effect without knowing any information other than the data itself. 
We have applied the method to all the K2 campaigns available at the Mikulski Archive for Space Telescopes, and we have tested the effectiveness of the procedure and its capability in preserving the astrophysical signal on a few transits and on eclipsing binaries.
 One {product} of this work is the identification of stable sources along the ecliptic plane that can be used as photometric calibrators for the upcoming Atmospheric Remote-sensing Exoplanet Large-survey mission.


\end{abstract}

\section{Introduction}

Accurate light curves photometric analysis often is limited by systematics. This is particularly crucial when systematics dominate the errors or when very small signals (as for planetary transits) are searched. Systematic effects are data modification that can arise from many different reasons, the change in time of the characteristics of a detector (eg. degradation of the CCD coating), or a failure on a component of the instrument could be examples.

Here we discuss the case of the K2 mission \citep{Howetal2014}. It uses the same spacecraft of the Kepler mission \citep{Boretal2010} that, after a failure on two reaction wheels, has lost the functionality of observing continuously the same portion of the sky. The K2 mission therefore entails a series of sequential observing ‘Campaigns’ of fields $\sim100\times100$ $deg^2$ distributed around the ecliptic plane, limited to a duration of approximately 80 days.  

Due to the telescope motion, the K2 original data are affected by strong systematic errors that lead to a reduction on the photometry precision of a factor 2-3 than the original Kepler mission.  
An automated procedure uses the presearch data conditioning algorithm \citep[][hereafter PDC]{Stumetal2014} built to remove the systematics. In {that} procedure, a discrete wavelet analysis is carried {out} to divide {each light curve into} channels of different time scale{s}. Then, each channel is corrected separately {by maximizing a Bayesian posterior probability distribution function whose priors are obtained from a cotrending basis vector generated by a set of highly correlated and quiet stars \citep{Smitetal2012}, and the corrected light curves are obtained by combining the channels again.}

However, the need of a higher photometric precision for a better exoplanet detection led different groups to develop their own procedure. Many procedures rely on the measurements of the position of the stars and correcting the intensity that correlates with the position. The latter can be evaluated through the center-of-light or the Gaussian fit to the stellar point spread function as described in \cite{VaneJon2014}, by deriving the astrometric solutions from the behavior of multiple stars as in \cite{Huaetal2015}, or by employing Gaussian Processes as in \cite{Aigetal2015,Aigetal2016}, and in \cite{Crosetal2015} (see also \cite{Lunetal2015} and \cite{Armetal2015} for other methods that rely on the stellar position). Other techniques avoid the measurement of the position as the pixel level decorrelation (developed for the Spitzer observations by \cite{Demetal2015} and adapted by \cite{Lugetal2016} to K2). In this technique, the intensities from the pixels are normalized by the total flux in the chosen aperture and then used as basis vectors for a linear least-squares fit to the aperture-summed flux.
 
In this work, {since we are interested in time related systematics}, we develop an alternative procedure that applies the Principal Component Analysis (hereafter PCA) on the intensity of the light curves of a given Campaign with the goal to remove systematic effects {in order to preserve} the astrophysical signal. {PCA has been already used to remove systematic errors in transiting exoplanet spectroscopy data \citep{Thatetal10,Zelletal14}, here we apply it to photometric data, in which } we assume that systematics does not depend on the position across the detector, {and,} therefore,  that their effect is the same for all the light curve in a given campaign.

This work was motivated by {the search of a set of standard stars in order to obtain high photometric stability (absolute and relative) for upcoming exoplanetary missions. This is the case of} ARIEL \citep[Atmospheric Remote-sensing Exoplanet Large-survey mission, ][]{Tinetal2018} that is aimed at investigating the atmospheres of several hundreds {of} planets orbiting distant stars in order to address the fundamental questions on how planetary systems form and evolve. The observational strategy of Ariel is based on differential spectroscopy measurements in and out the transit (or occultation). Furthermore in several cases it will perform repeated visits on the same target to reach the needed signal to noise. As a consequence high photometric stability (absolute and relative) is a requirement in order to guarantee the capability to recover the planetary signal. Therefore it will be necessary to select a set of standard stars, as constant as possible, possibly spread out in various sky directions (in order to be sure that ARIEL may observe one of them at any time). The photometric standard stars normally used from ground do not guarantee to be enough stable, since from ground the stability requirement is typically of the order of 0.01, while ARIEL will need stability of about 1-2 order of magnitude higher. On the contrary modern photometric missions from space are able to reach the needed precision, after removal of instrumental effects {\citep[ex. Kepler,][]{Boretal2010}}. Among the missions that achieve large fields of view (CoRoT, Kepler and K2) K2 is the most suitable for this purpose because may observe in a area spread on several sky directions.

 We describe the data preparation in Section~\ref{sec:data}. In Section~\ref{sec:met} we present the method while in Section~\ref{sec:res} we present the results, and we discuss them in Section~\ref{sec:disc}.

\section{Data preparation}
\label{sec:data}

Our analysis has been conducted using all the K2 Data available at the MAST Archive by June 2019. We have selected 'stars' (Object type) with calibrated 'long cadence' (Target type) light curves (see Tab.~\ref{tab:Camps}), with a magnitude brighter than 11 (Kepler Magnitude), i.e. stars sufficiently bright to have a high signal-to-noise ratio in K2, and with an observing time between 75 and 84 days. {Selected} data from each Campaign have been divided in two data set{s}, those brighter than 10 (dataset A) and the others (dataset B). We have applied the analysis on dataset A and checked its robustness by extendin{g} it to dataset B (see Section~\ref{sec:stbsour}).
Data are available both original (raw) and PDC corrected. Single randomly distributed NAN are present in both as well as for long range intervals (few days). {We have completely removed them from the light curves, therefore our procedure has been applied to the subset of the raw data in which the temporal bins of all sources (common between all the light curves of a campaign) have finite values. This procedure excludes from the analysis at most few percent of the time.}

\begin{table}
\caption{Number of light curves in each campaign, and in two different magnitude ranges. {(Campaign 10 and 11 are divided in two data set. In the case of Campaign 10, only the second set were available while for Campaign 11 both, and they have been analysed separately)}.}
  \label{tab:Camps}
\begin{tabular}{ lll } 
\hline\noalign{\smallskip}
     Campaign    & A & B  \\ 
\noalign{\smallskip}\hline\noalign{\smallskip}

	  C01     &  $338$  &  $549$       \\ 
 	  C02     &  $601$  &  $676$       \\ 
      C03     &  $407$  &  $758$       \\ 
      C04     &  $689$  &   $1018$    \\ 
      C05     &  $793$  &  $1135$       \\ 
      C06     &  $536$  &  $778$      \\ 
      C07     &  $405$  &  $448$      \\ 
      C08     &  $538$  &  $687$      \\ 
      C10     &  $261$  &   $696$     \\ 
      C11     &   $1535$ &   $2782$   \\ 
      C12     &  $463$  &   $717$      \\ 
      C13     &  $568$  &   $861$      \\ 
      C14     &  $461$  &   $657$      \\ 
      C15     &  $645$  &   $1171$     \\ 
      C16     &  $655$  &   $1018$     \\ 
      C17     &  $497$  &   $688$     \\ 
      C18     &  $543$  &  $798$       \\ 
      C19     &  $516$  &  $787$     \\ 
\noalign{\smallskip}\hline\noalign{\smallskip}
      Total   &  $10451$  &  $16224$     \\ 
\hline\noalign{\smallskip}
\end{tabular}

\end{table}

\section{The method}
 \label{sec:met}
The Principal Component Analysis (hereafter PCA) is a statistical technique that uses an orthogonal linear transformation to convert a set of correlated variables into an uncorrelated one \cite[see][for an implementation in python]{Rascetal17}. It performs a change of coordinate system such that variables are described along directions that maximize the correlation between them, with the highly correlated components explaining the most of the variance of the set:

\begin{equation}
 Y = X A
\label{eq:pca}
\end{equation}

\noindent where $ {X}$ is a set of data to be analysed, the matrix $A$ contains the coefficients of the transformation, and $ {Y}$ is the set of the transformed data. The matrix of coefficients $A$ is made by eigenvectors of the covariance matrix, i.e. the principal components, {whose elements are of the form $CovMat(i,j) = \sum_{k=1}^p(x_{ik}-\bar{x}_i)(x_{kj}-\bar{x}_j)/(p-1)$, where $p$ is the number of rows in the data set, $x_{ik}$ and $x_{kj}$ are data elements, $\bar{x}_i$ and $\bar{x}_j$ are, respectively, the mean of the $i$th and $j$th column of the $X$ matrix. The covariance matrix and its eigenvectors satisfy} the equation

\begin{equation}
CovMat(X) A_i= \lambda_i A_i
\label{eq:covmat}
\end{equation}     

\noindent where $\lambda_i$ is the eigenvalue of the eigenvector $A_i$, from which the relative variance of the data set along the direction of the component ($var_i$) can be computed as $var_i = \lambda_i/(\sum_i \lambda_i)$.

Once the principal components and the transformed data are computed, the set of data $X'$ can be reconstructed back by reversing eq.~\ref{eq:pca} as follow{s}

\begin{equation}
X' = Y A^{t}
\label{eq:recx}
\end{equation}

\noindent where we used the orthogonality and the normalization of the eigenvectors, i.e. $A^t A=I$, with $A^t$ is the transposed of $A$ and $I$ is the identity matrix.

 Since we assume that, for a given campaign, systematic effects affect all the light curves in the same way, we expect light curves to be highly correlated with only few components explaining most of the variance (due to the systematics) and the remaining ones explaining the very small variance due to the uncorrelated intrinsic astrophysical signal among with the statistical noise.
Since eigenvalues and, accordingly, eigenvectors are in decreasing order {of the explained variance}, we can reconstruct the {''true''} data set $X$ with eq.~\ref{eq:recx} by removing the first principal components, related to the systematics, and taking only the components with low variance, related to the astrophysical signal.  {Of course, if we eliminate too many components, we will remove part of the astrophysical signal, therefore, we have to find the principal component that separates the region in which eigenvectors describe the systematics by the region in which they describe the astrophysical signal. We will refer to this component as the cut-off component {(see Section~\ref{subsec:pca2lc}). We will determine it by analysing how the variance varies when we iteratively remove the component with the highest variance. The idea is that the variance explained by components due to uncorrelated signals is low while the one due to highly correlated components is high, therefore, since we will remove iteratively one component at time, we will see a strong variation in the explained variance until we have removed most of the correlated signal due to systematics. The cut-off component will be the one since the variation of the variance become smaller.}

To account for the many different amplitudes of the stellar variability, {we have standardized each light curves (i.e. a column of the X data) shifting it by its mean and dividing by its standard deviation. Consequently, the covariance matrix becomes the correlation matrix, whose elements are of the form $CorMat(i,j) = \sum_{k=1}^p((x_{ik}-\bar{x}_i)/\sigma_i)(x_{kj}-\bar{x}_j)/\sigma_j))/(p-1)$, where $\sigma_i$ and $\sigma_j$ are the standard deviation, respectively, of the $i$th and $j$ column of matrix $X$.}

\subsection{PCA applied to light curves}
\label{subsec:pca2lc}
In this study, the PCA is applied separately to each Campaign described in Section~\ref{sec:data} for which a set of data $X$ is a $(p \times n)$ matrix, where $n$ is the number of light curves (see Tab.~\ref{tab:Camps}) while $p$ is the number of temporal bins used to study each light curve (ranging from $\sim2000$ to $\sim4000$). The matrix of coefficients $A$ is a $(n \times n)$ matrix (as well as for $CorMat$) and is made by (column) eigenvectors $(n \times 1)$, which are related to $n$ eigenvalues.  

Since we want to eliminate the correlated signal that comes from the first components but we do not know the cut-off component, we run iteratively the PCA by removing one component at time. We expect that at the first iteration {we remove the most correlated component and, therefore,} the eigenvalues distribution is highly varying but it becomes flatter after at each iteration, converging to the distribution of a completely uncorrelated set of data. {As reference of the limit of an uncorrelated set of data, we obtain its distribution by} simulating a new set of data by applying a bootstrap on the light curves, i.e. by randomizing the temporal bins. We set a maximum of 80 iteration{s. This number should be compared with the total number of eigenvectors, that is the number of the light curves in a campaign (of the order of hundreds), rather than the number of temporal bins (of the order of thousands), therefore by using a maximum of 80 iterations we are conservative since we have verified that, removing 80 components, a substantial fraction of the astrophysical signal is removed (as verified in light curves with transiting planets). In fact the cut-off component is always much smaller than 80.}

 We then compare the variances distribution obtained from each iteration {with the previous one}. At each iteration, the entire set of data is reconstructed (from eq.~\ref{eq:recx}) in order to feed the PCA at the subsequent iteration with the new set of data, reconstructed without the component carrying the greatest variance. 

\begin{figure*}
	\centering
	\resizebox{\textwidth}{!}
	{
	\includegraphics[scale=1]{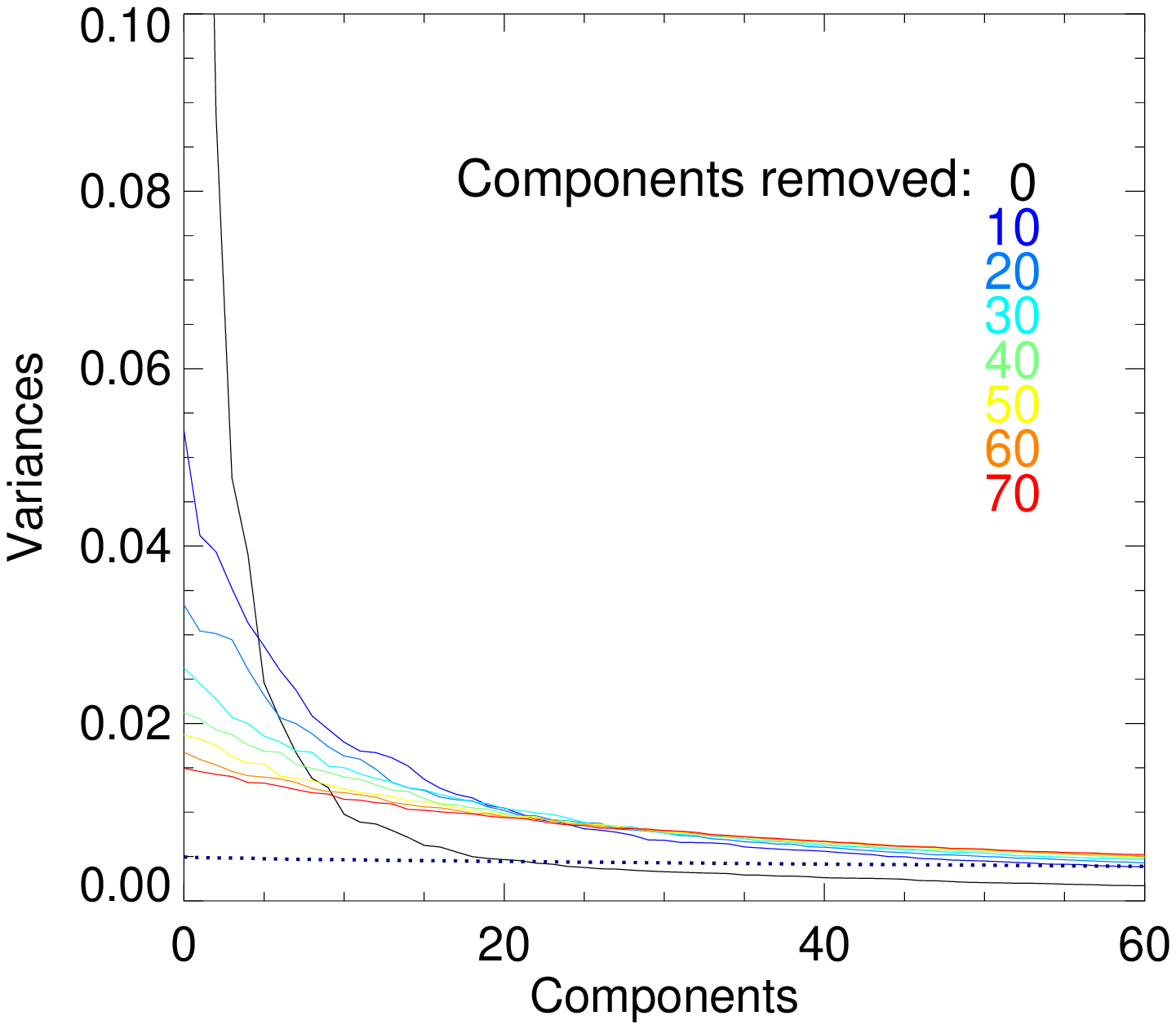}
	\includegraphics[scale=1]{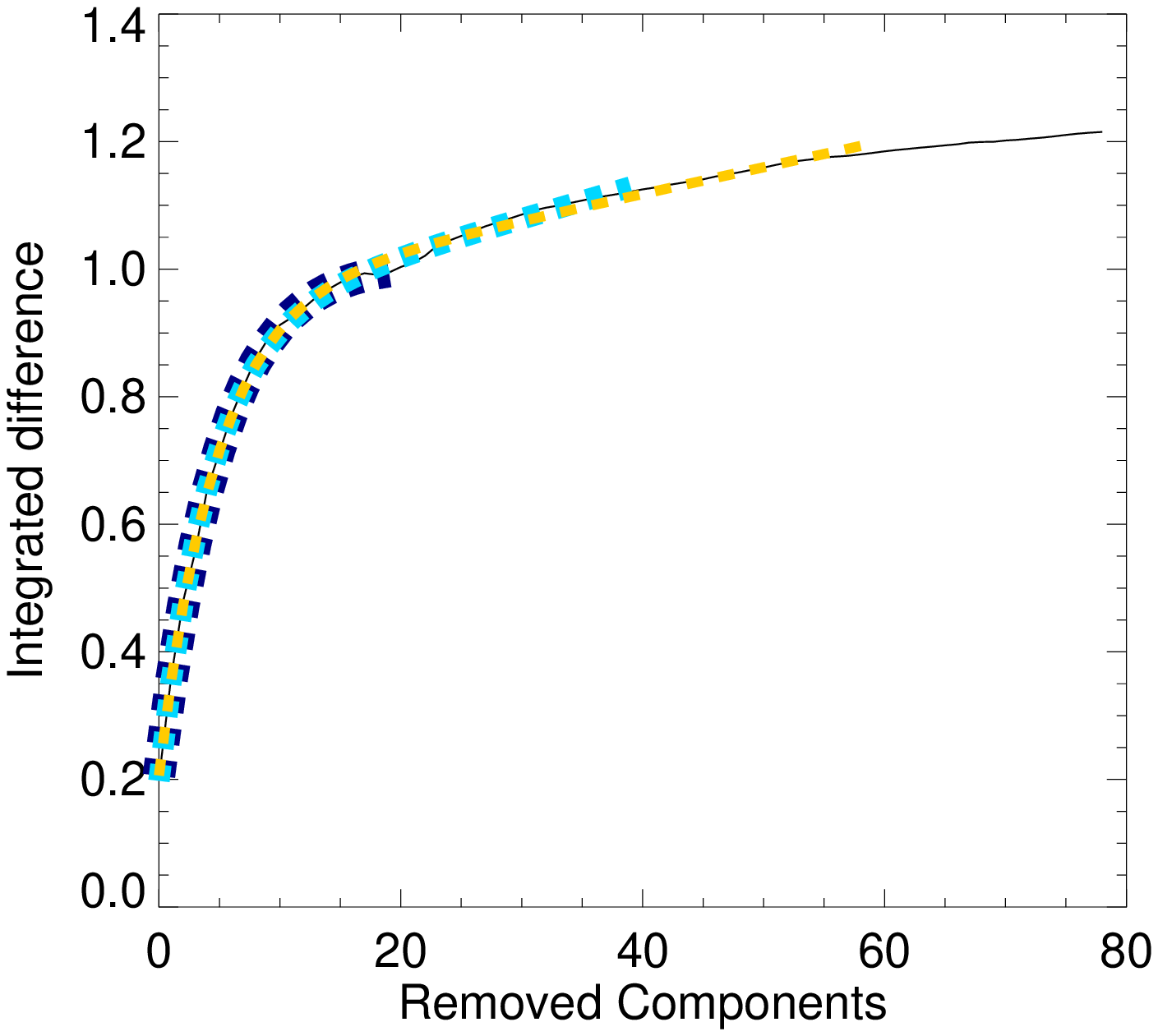}
	}
\caption{(Left) Variances distributions obtained by applying the PCA on the original data, with a set of removed components up to 60 for visual reasons. The distributions are colour coded based on the number of components removed. We also show (dashed line) the variance distribution of a simulated completely uncorrelated set of data. (Right) Integrated difference (as defined in eq.~\ref{eq:intdiff}) at different components removed, in which we show the fit for 20,40,60 components, color coded as for the left figure.}
\label{fig:evals}
\end{figure*}

In Fig.~\ref{fig:evals}(left) we show the variance explained by removing multiples of ten components, in the case of Campaign 1. As one could expect the variance rapidly fall{s} after removing the first components, however the distribution of the variance does not converge to an equally distributed variance distribution (dashed line in Fig.~\ref{fig:evals}(left)). This could be due to correlation at small temporal scales of astrophysical signals in the light curves or non linear effects, and it is a common behaviour among all the campaigns.

To estimate the cut-off, we study the variation of variances distribution at each iteration. We expect that the variation of the variance distribution, with respect the first iteration, depends on the nature of the component we are removing, if it is a systematics the variation will be higher than the variation obtained by removing the astrophysical signal because systematics are {shared between all the light curves and therefore they produce} highly correlated {signal} while {amplitude and timing of} the astrophysical signal is {different from one source to the others and it produces an uncorrelated signal, therefore} we expect a different regime of variation before and after the cut-off component. To quantify the difference of variance distributions, we evaluate at each iteration the integrated difference (hereafter IntDiff), expressed by

\begin{equation}
IntDiff[j] = \sum_i |var[i,j]-var[i,0]|
\label{eq:intdiff}
\end{equation} 
\noindent where $var$ is the variance, indexes $i$ and $j$ indicate, respectively, the component and the iteration. The result of this procedure for Campaign 1 is presented in Fig.~\ref{fig:evals}(right). The curve can be divided in three parts, the first is characterized by a rapid growth, followed by a quasi-linear part and then a flat region. 
{Our interpretation is that the} first part is dominated by the systematic effects, that we want to remove, while the others by astrophysical signal and noise. Now, we want to derive the component that separates the first two parts (i.e. the cut-off component), therefore we fit the IntDiff curve with the function $A+\vert \textbf{B} \vert x+C e^{-x/\tau}$ that is able to describe the initial rapid (exponential) growth and the quasi-linear part. 
 
To describe the first two parts as best as we can, we repeated the fit starting from the range of (removed) components [0,20] and iteratively increasing the range one component at time, up to [0,80]. For each iteration, the best fit has been chosen minimizing the $\chi^2$, then we compare all the best fit obtained from each iteration and we take as a final best fit the one which has the minimum $\chi^2$ between iterations. The final best fit, for all the campaigns, is presented in Fig.~\ref{fig:comps} and Fig.~\ref{fig:comps1}.

We consider  {the integer of} $3\tau$ as a good choice for the cut-off because it reduces the exponential part by $\sim95$\% and it, effectively, separates the exponential range from the linear region. {In Fig.~\ref{fig:comps1} we also show the obtained cut-off component as a function of the Campaign. The choice on the $\tau$ parameter leads to values in the range $6-26$, with the majority of the cut-offs lying in the range $11-16$ and only $4$ values outside.}

\begin{figure*}[!th]
	\centering
	\resizebox{\textwidth}{!}
	{
	\includegraphics[width=0.39\columnwidth ]{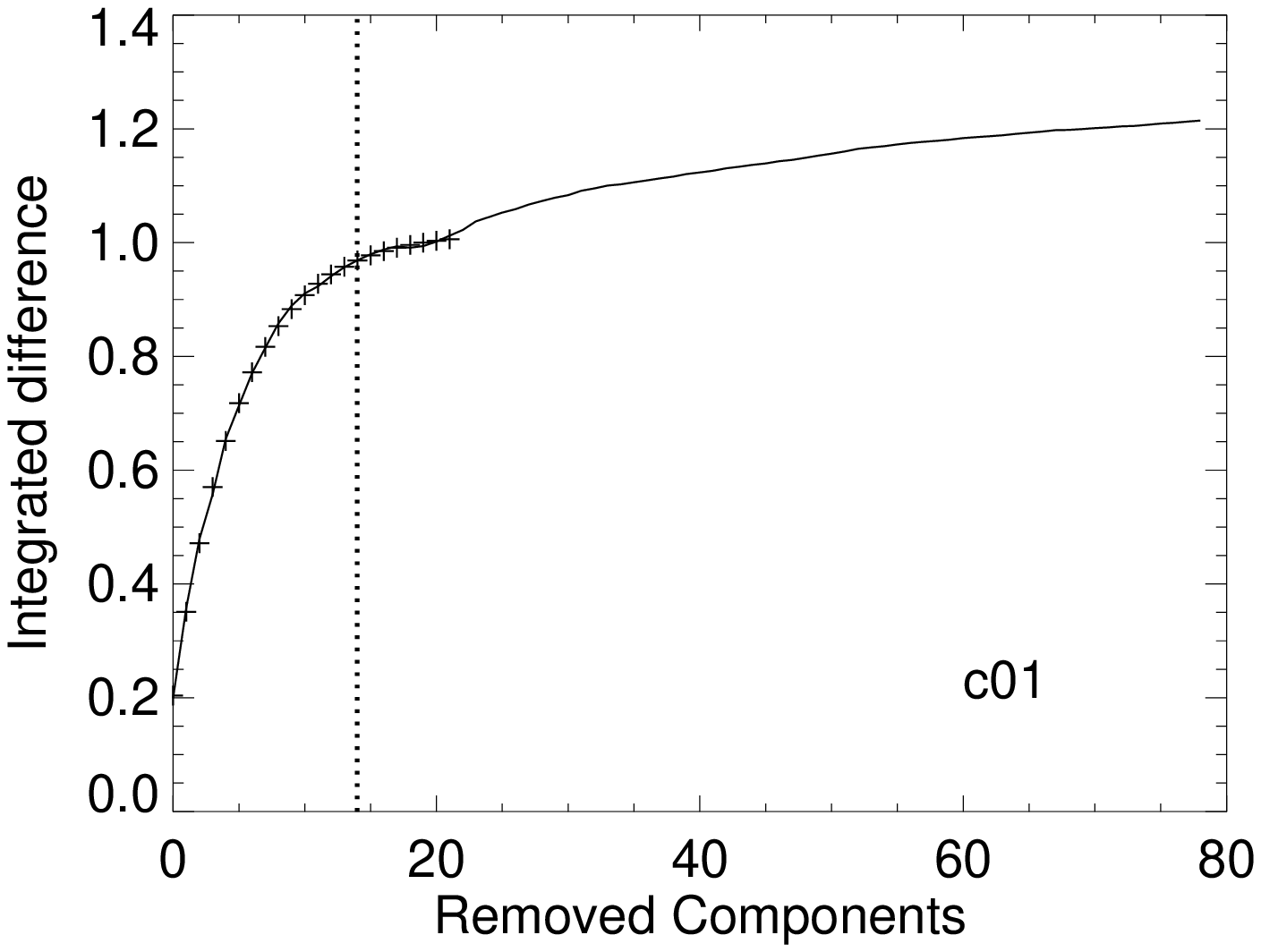}  
	\includegraphics[width=0.39\columnwidth ]{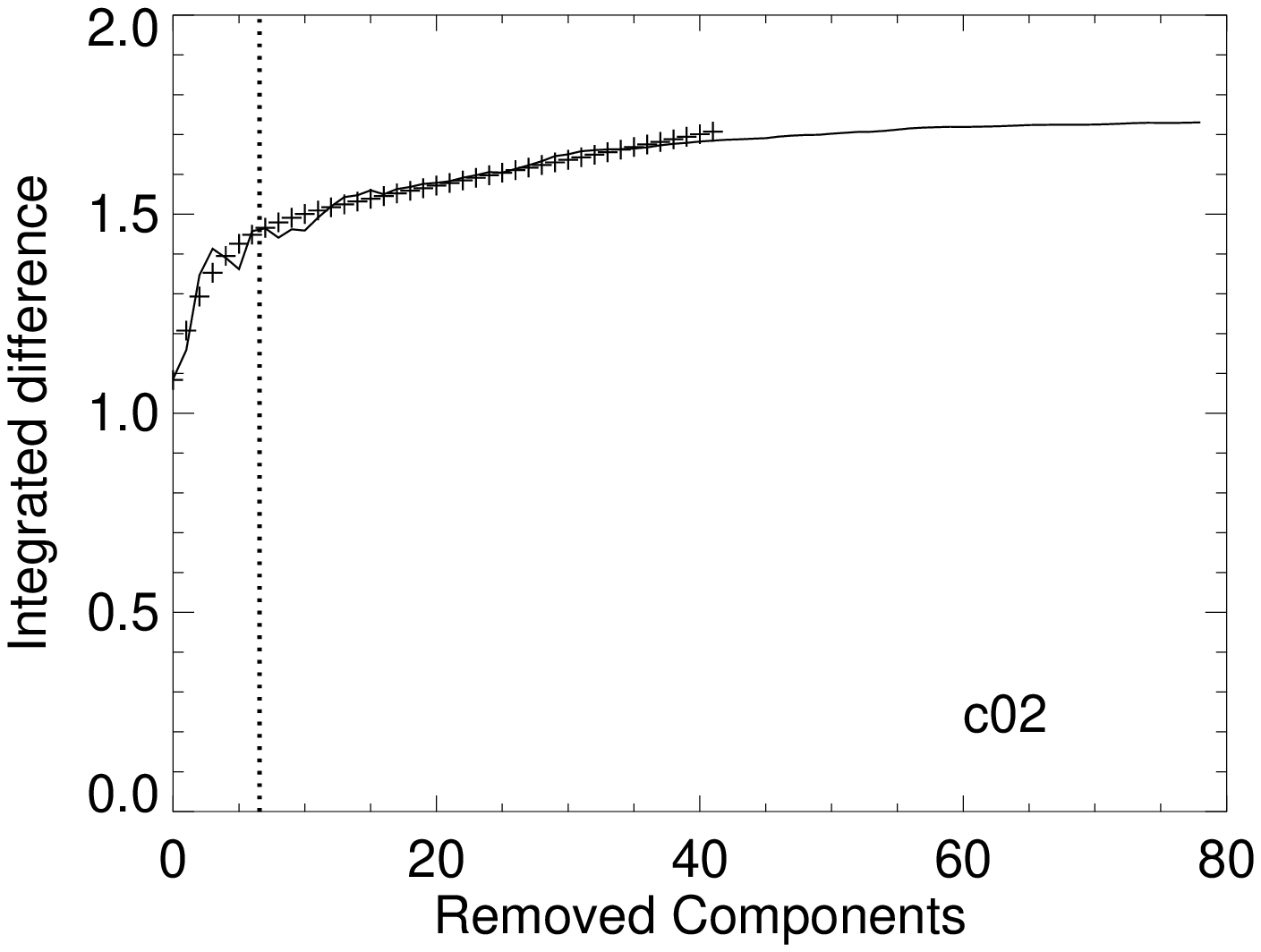}

	}
	\resizebox{\textwidth}{!}
	{
	\includegraphics[width=0.39\columnwidth ]{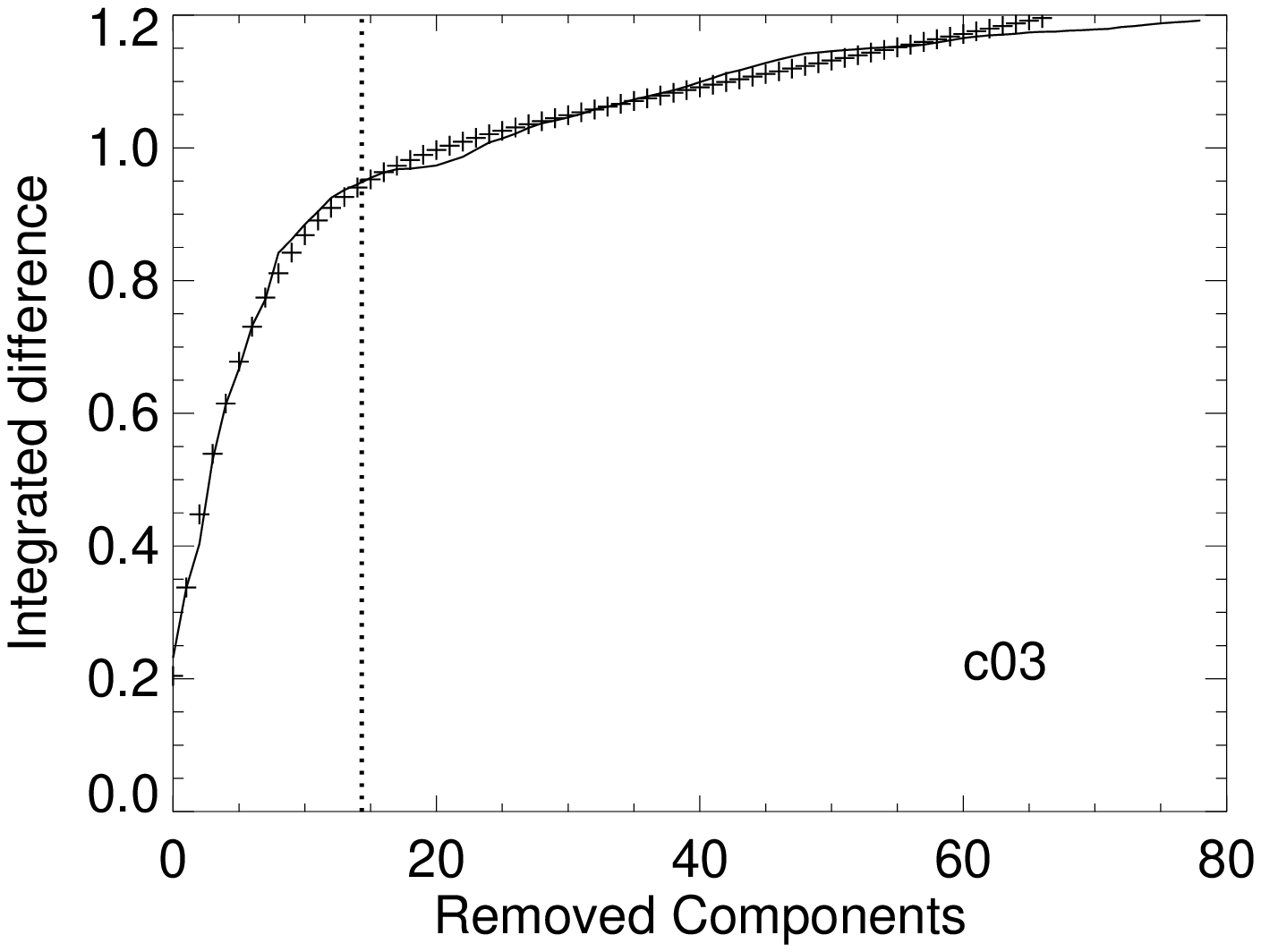}
	\includegraphics[width=0.39\columnwidth ]{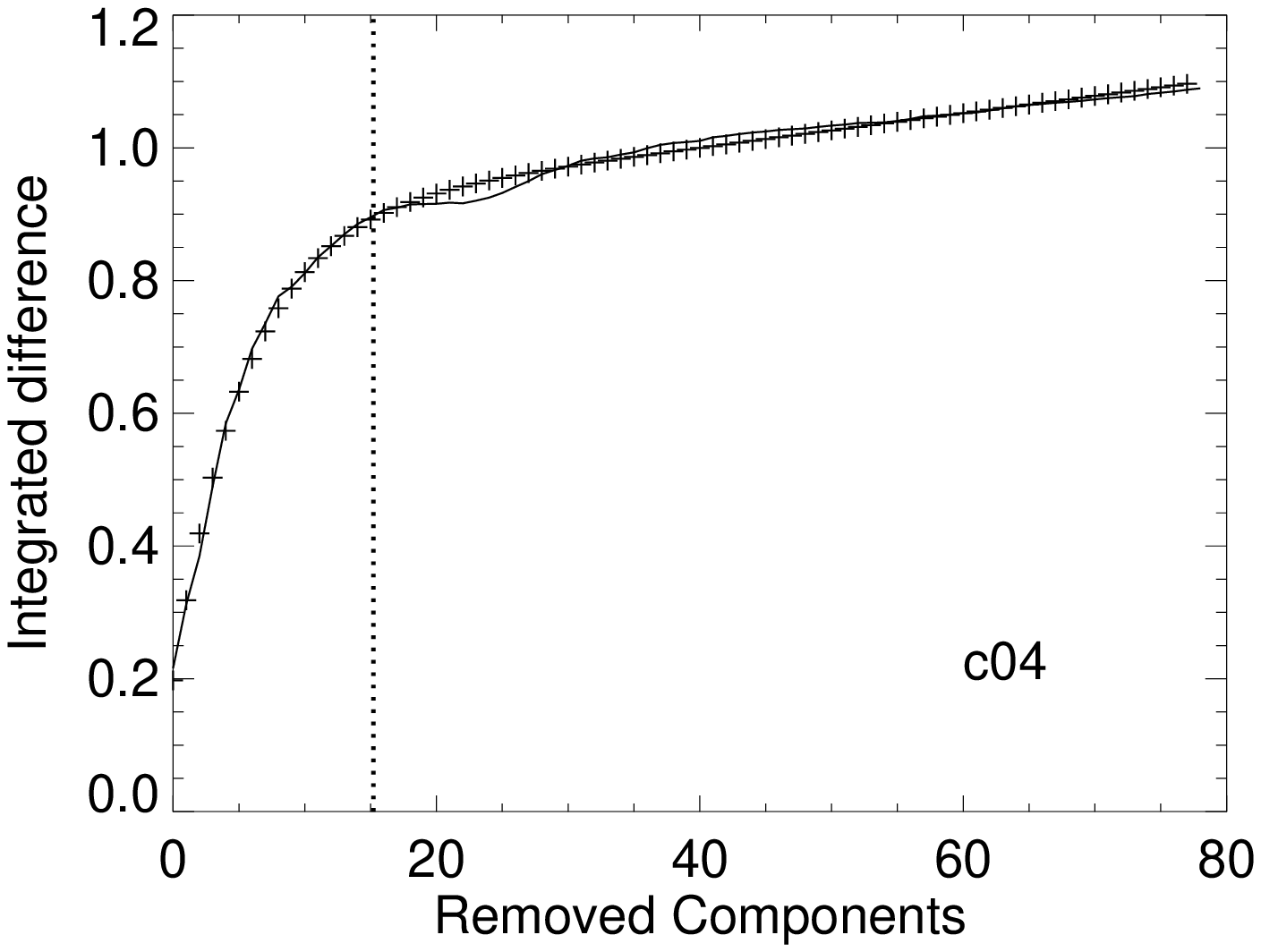}  
	}
	\resizebox{\textwidth}{!}
	{
	\includegraphics[width=0.39\columnwidth ]{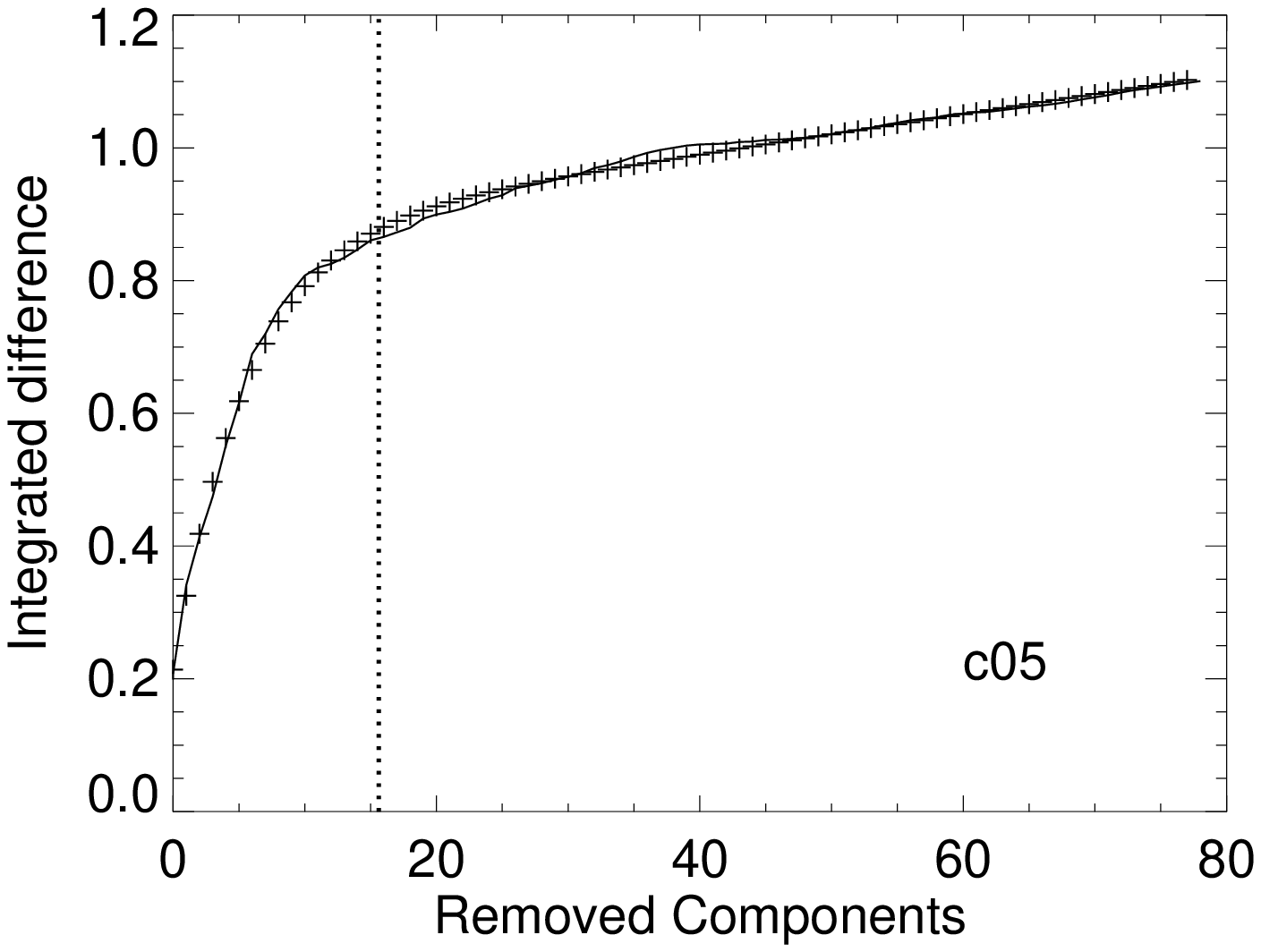}
	\includegraphics[width=0.39\columnwidth ]{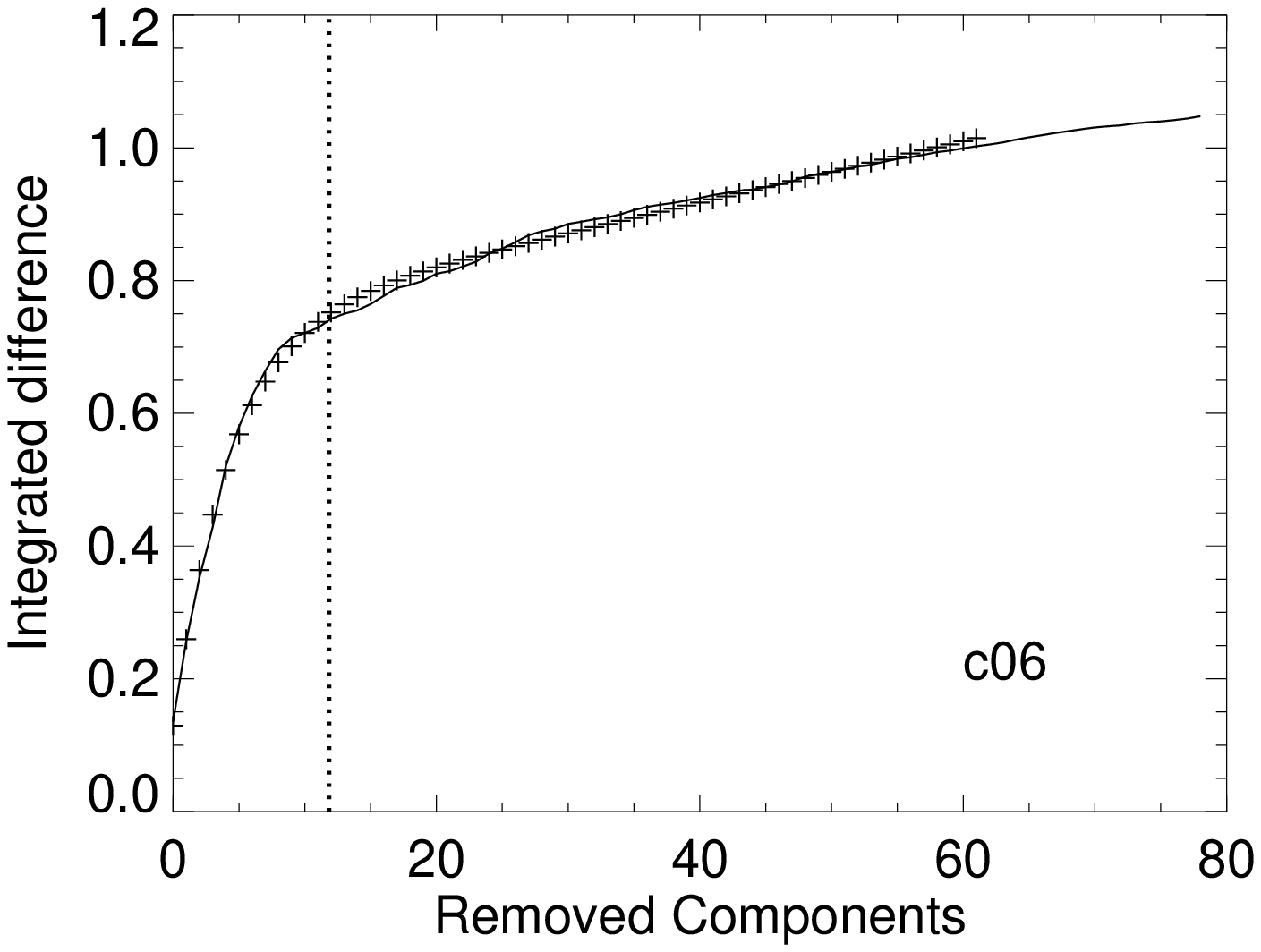}  
	}
	\resizebox{\textwidth}{!}
	{
	\includegraphics[width=0.39\columnwidth ]{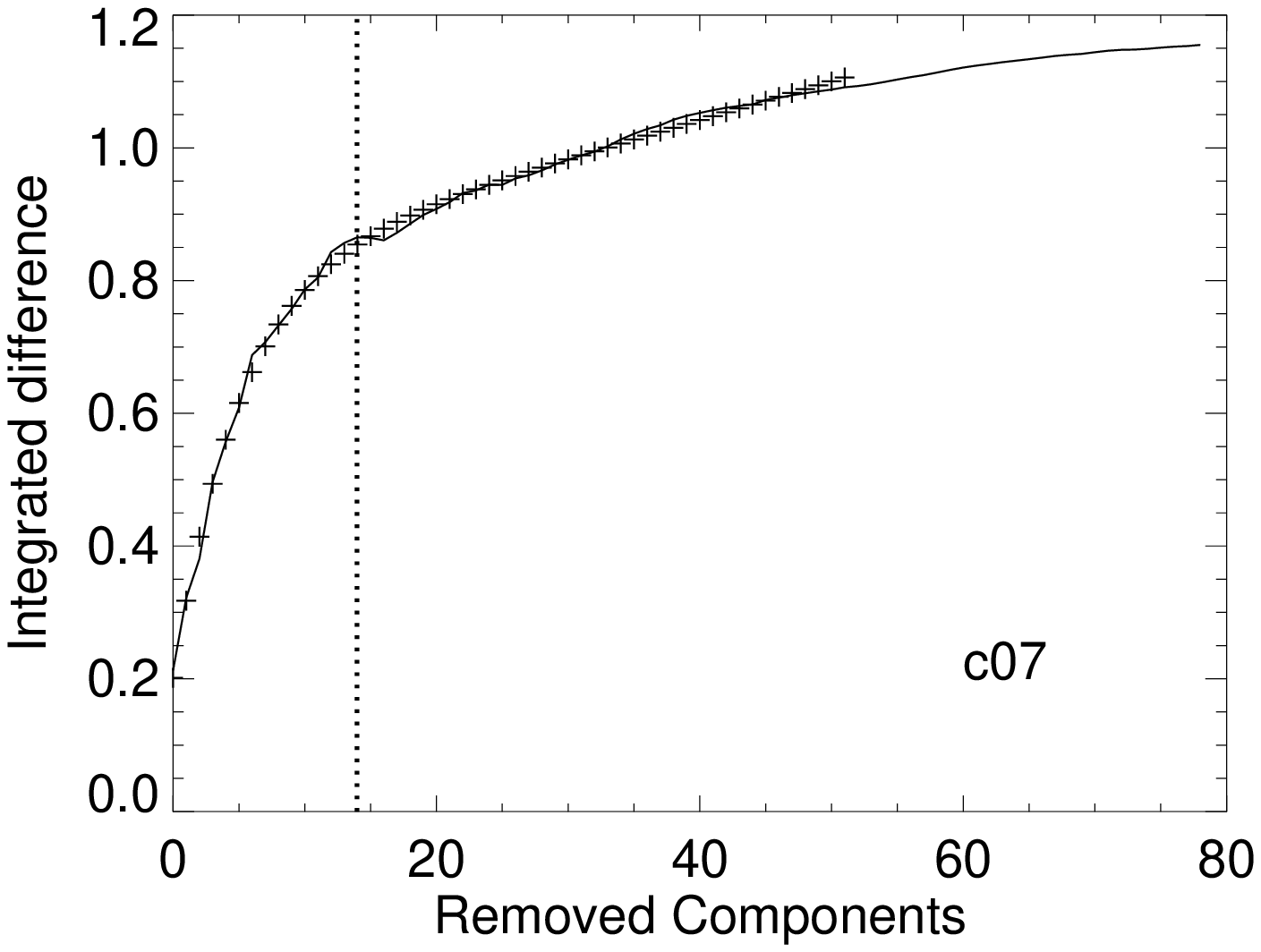}
	\includegraphics[width=0.39\columnwidth ]{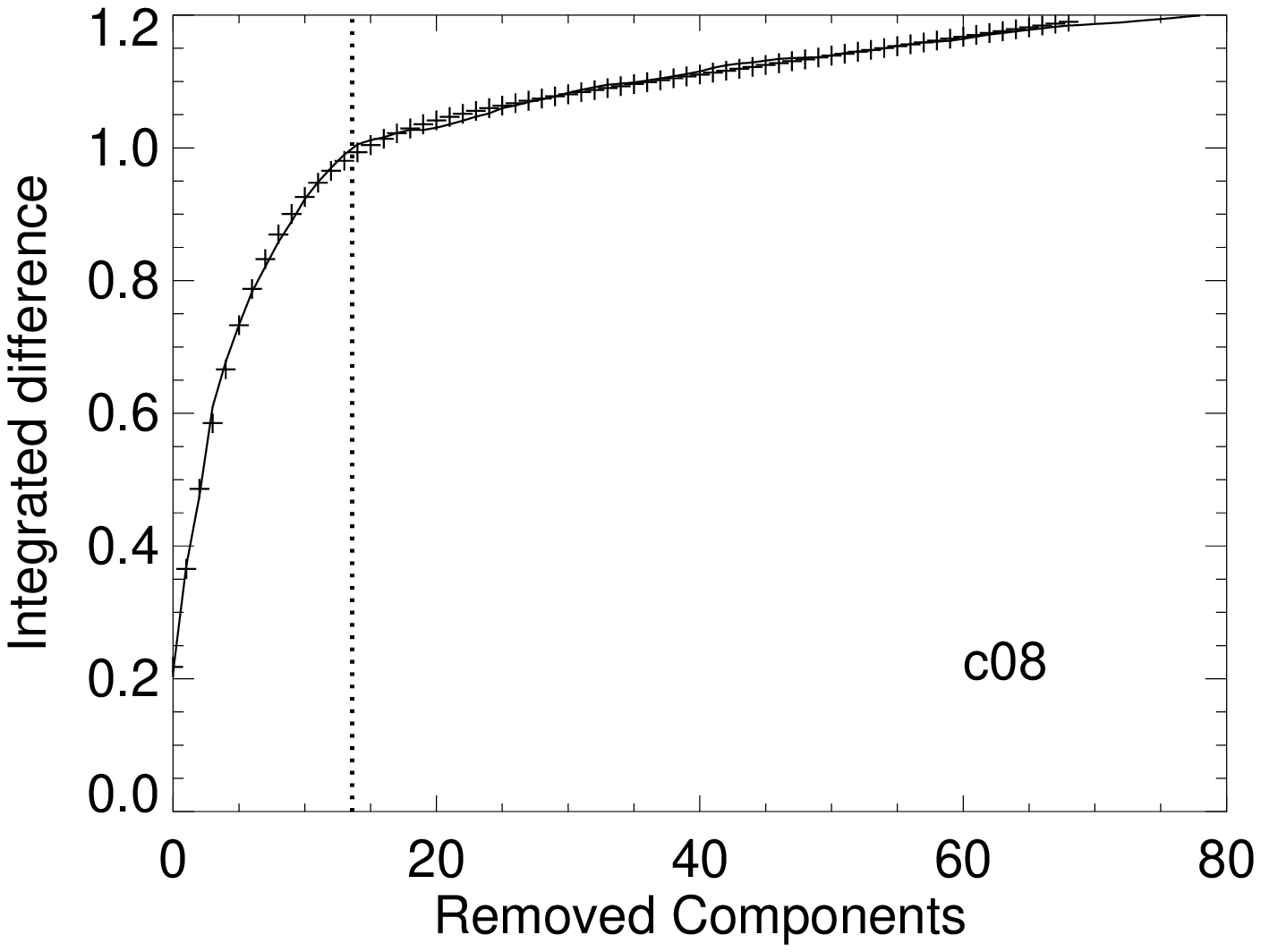}  
	}
	\resizebox{\textwidth}{!}
	{
	\includegraphics[width=0.39\columnwidth ]{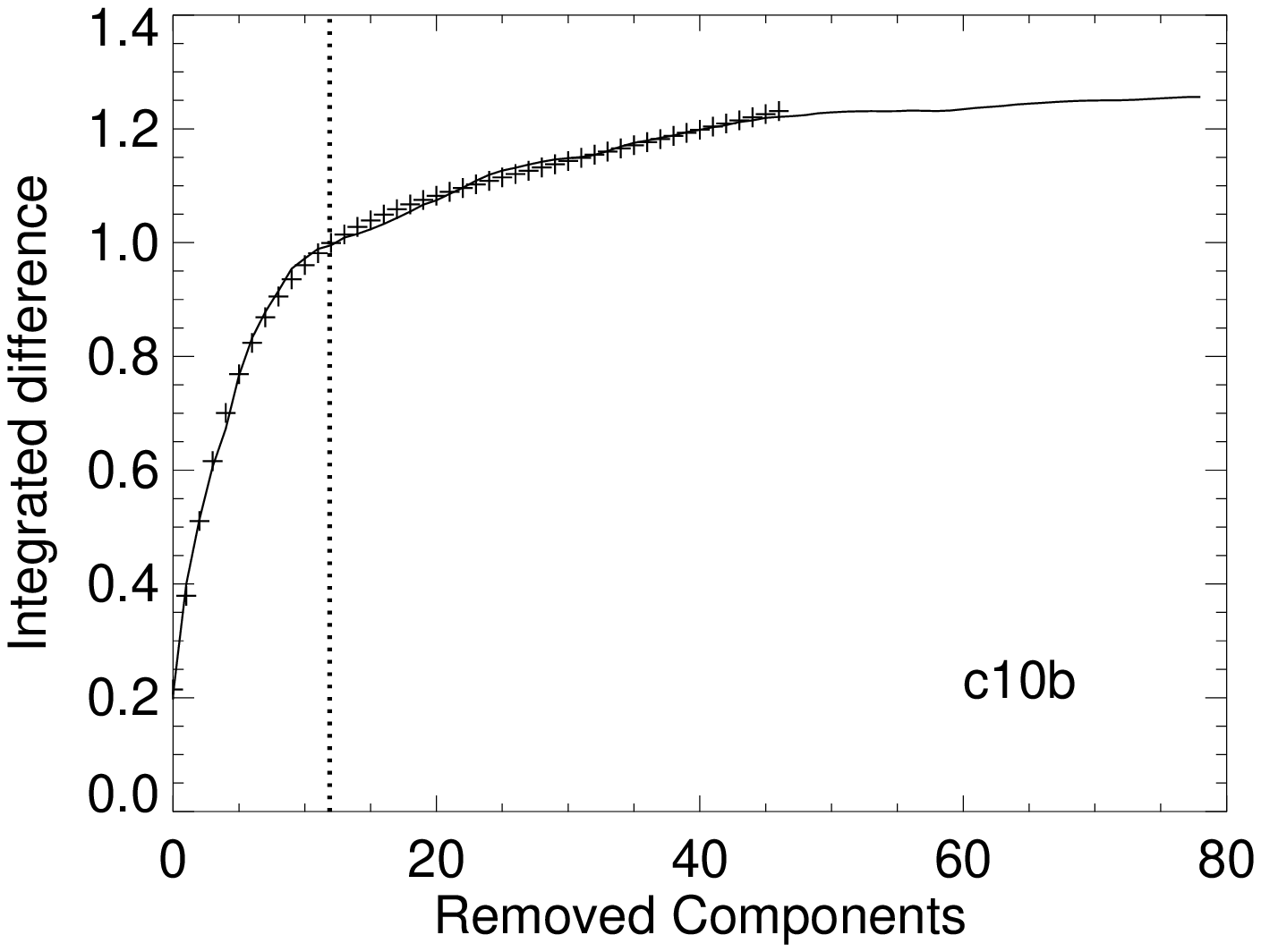}
	\includegraphics[width=0.39\columnwidth ]{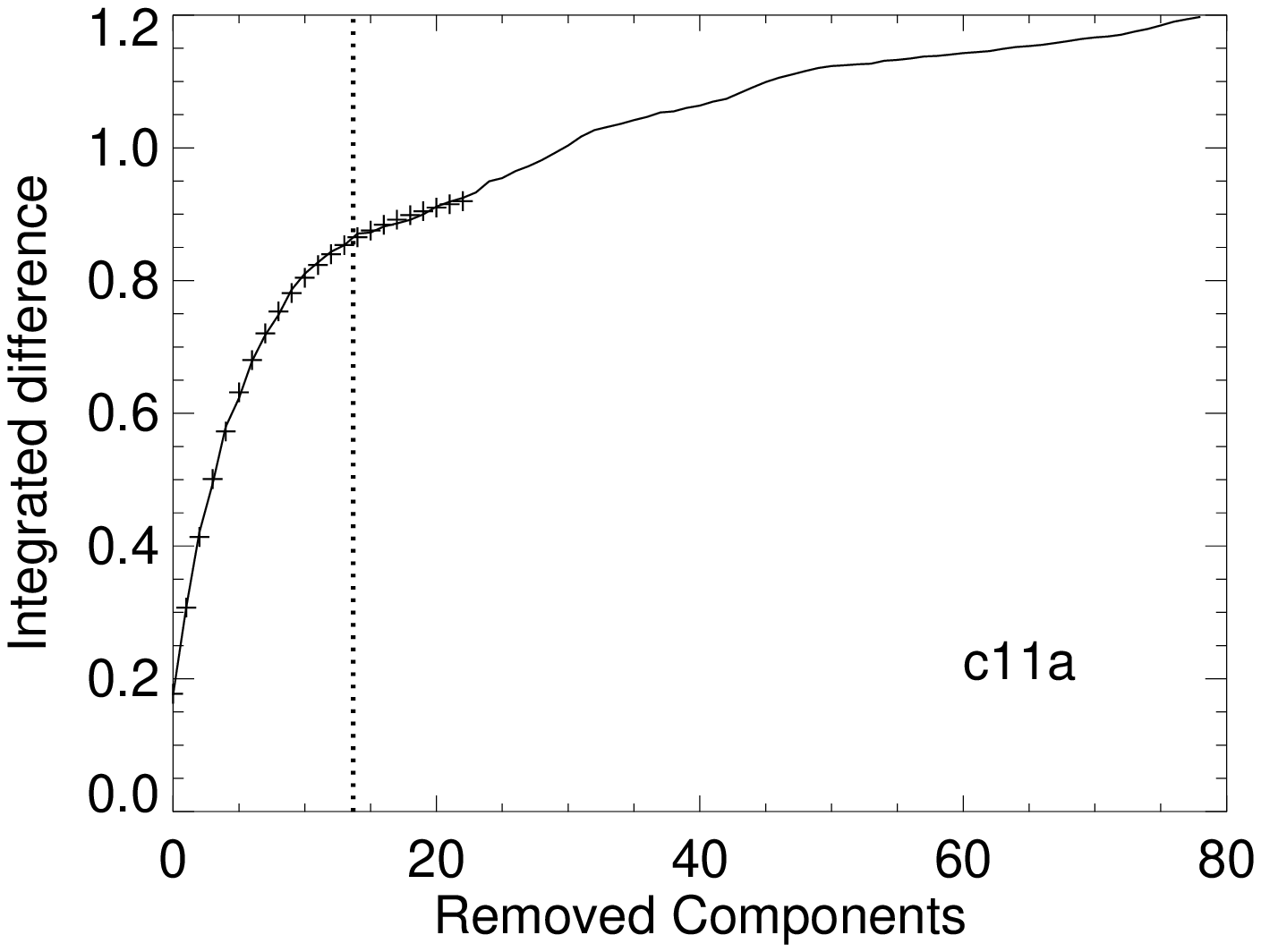}  
	}
\caption{Integrated difference of the variance at different component removed (black line) for Campaign 1 to 11a, in which the best fit (cross points) and the component at $3\tau$ (vertical dashed line) are shown.}
\label{fig:comps}
	\end{figure*}

\begin{figure*}[!th]
	\centering
	\resizebox{\textwidth}{!}
	{
	\includegraphics[width=0.39\columnwidth ]{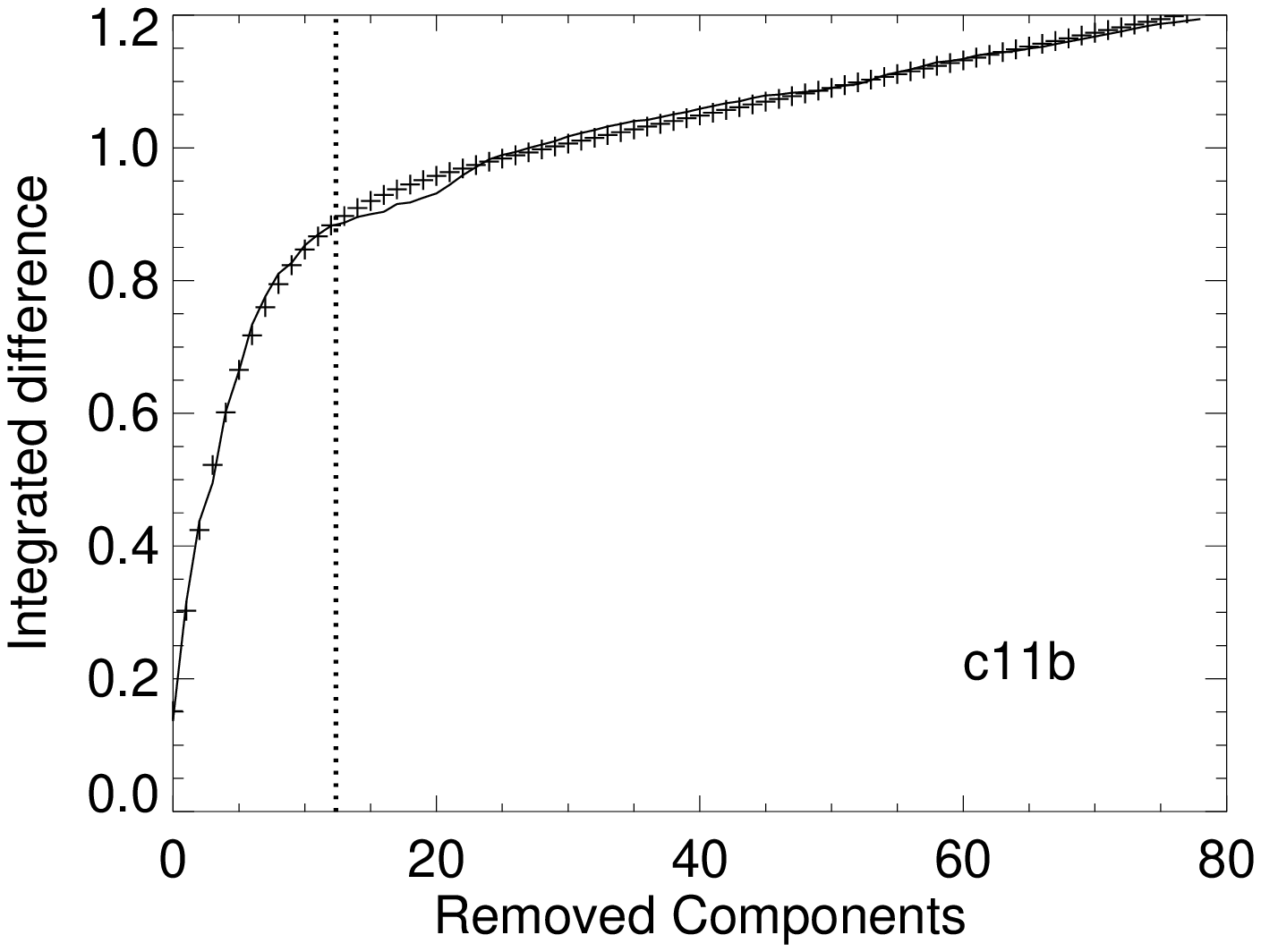}  
	\includegraphics[width=0.39\columnwidth ]{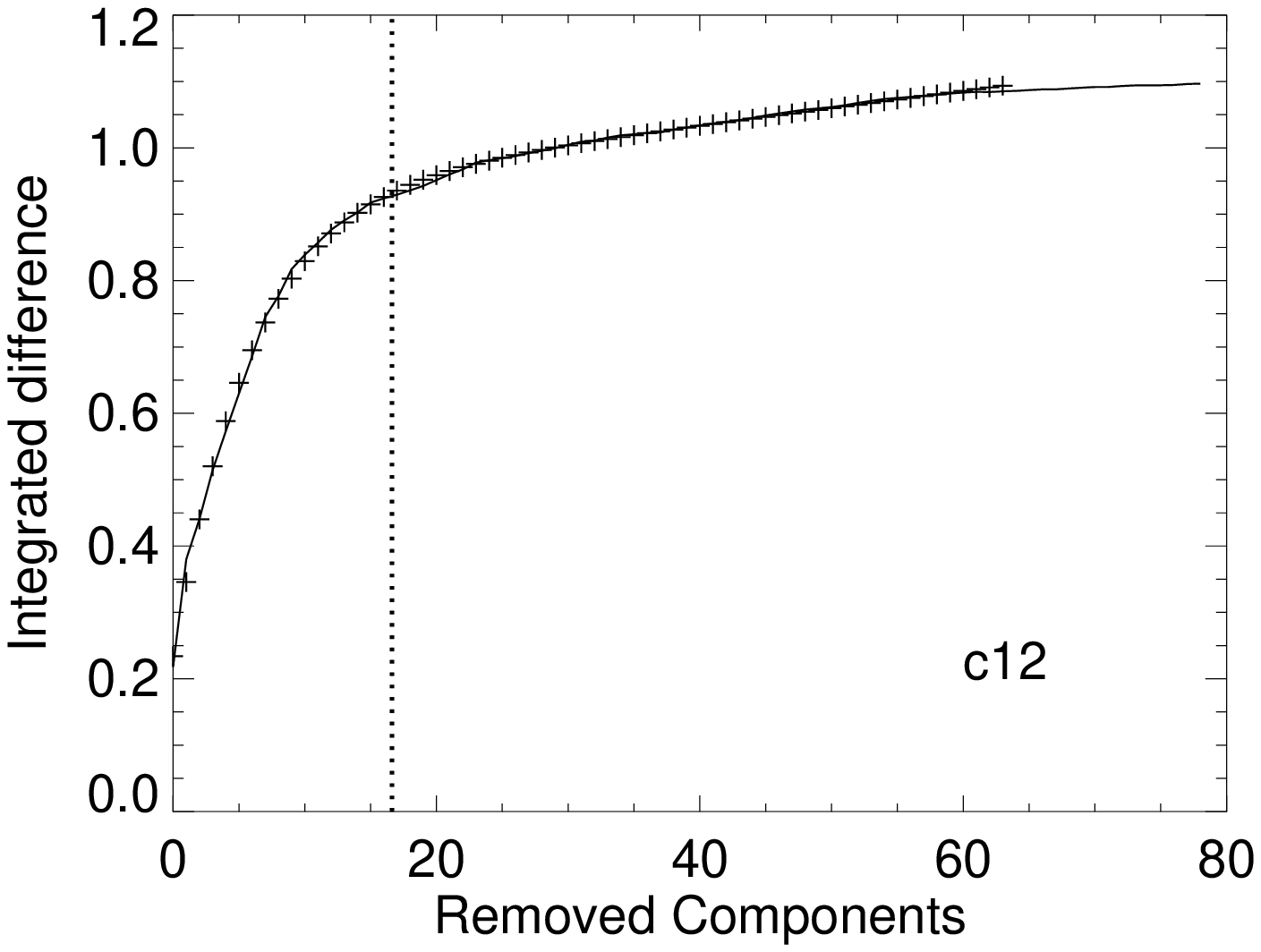}

	}
	\resizebox{\textwidth}{!}
	{
	\includegraphics[width=0.39\columnwidth ]{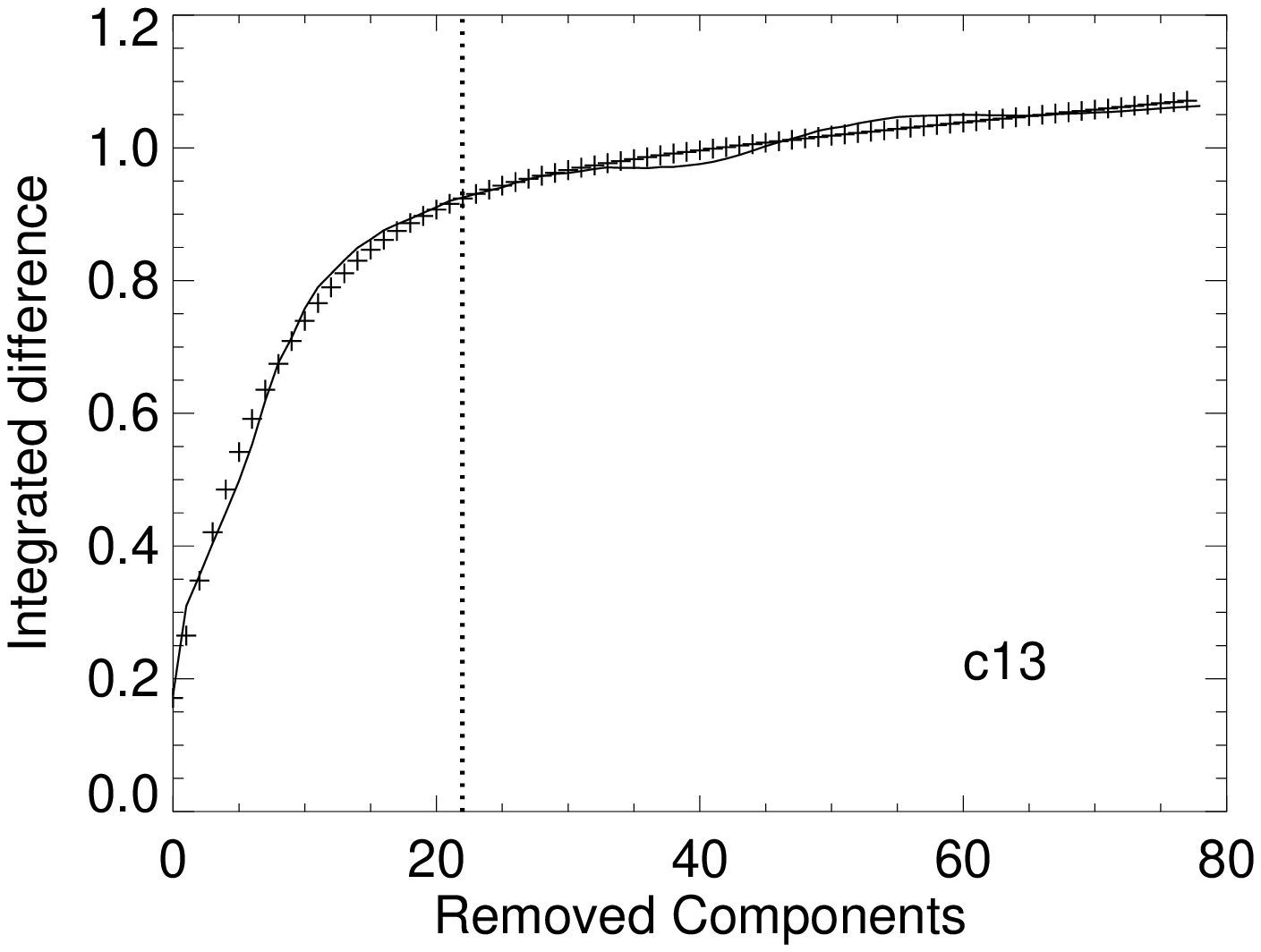}
	\includegraphics[width=0.39\columnwidth ]{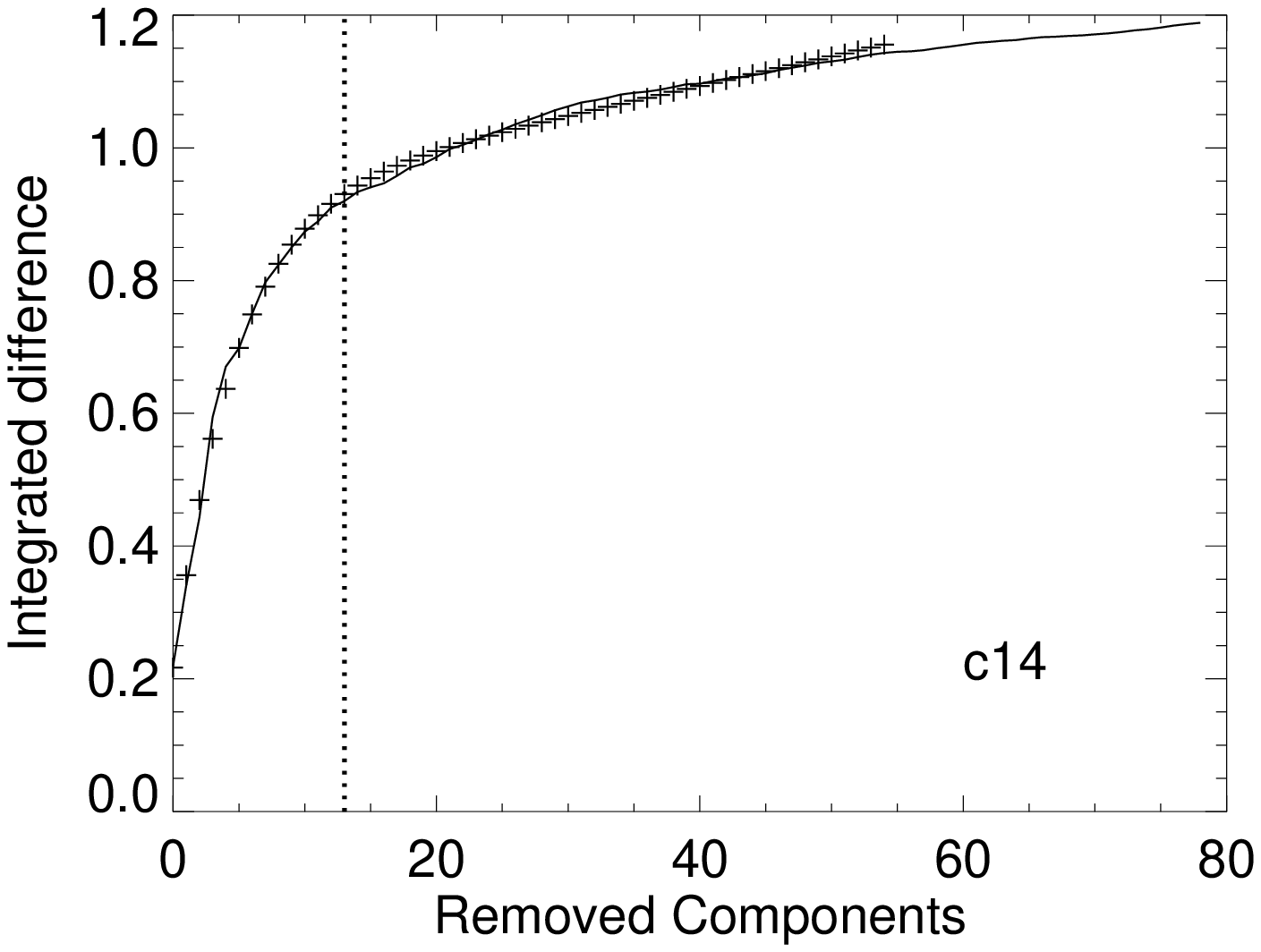}  
	}
	\resizebox{\textwidth}{!}
	{
	\includegraphics[width=0.39\columnwidth ]{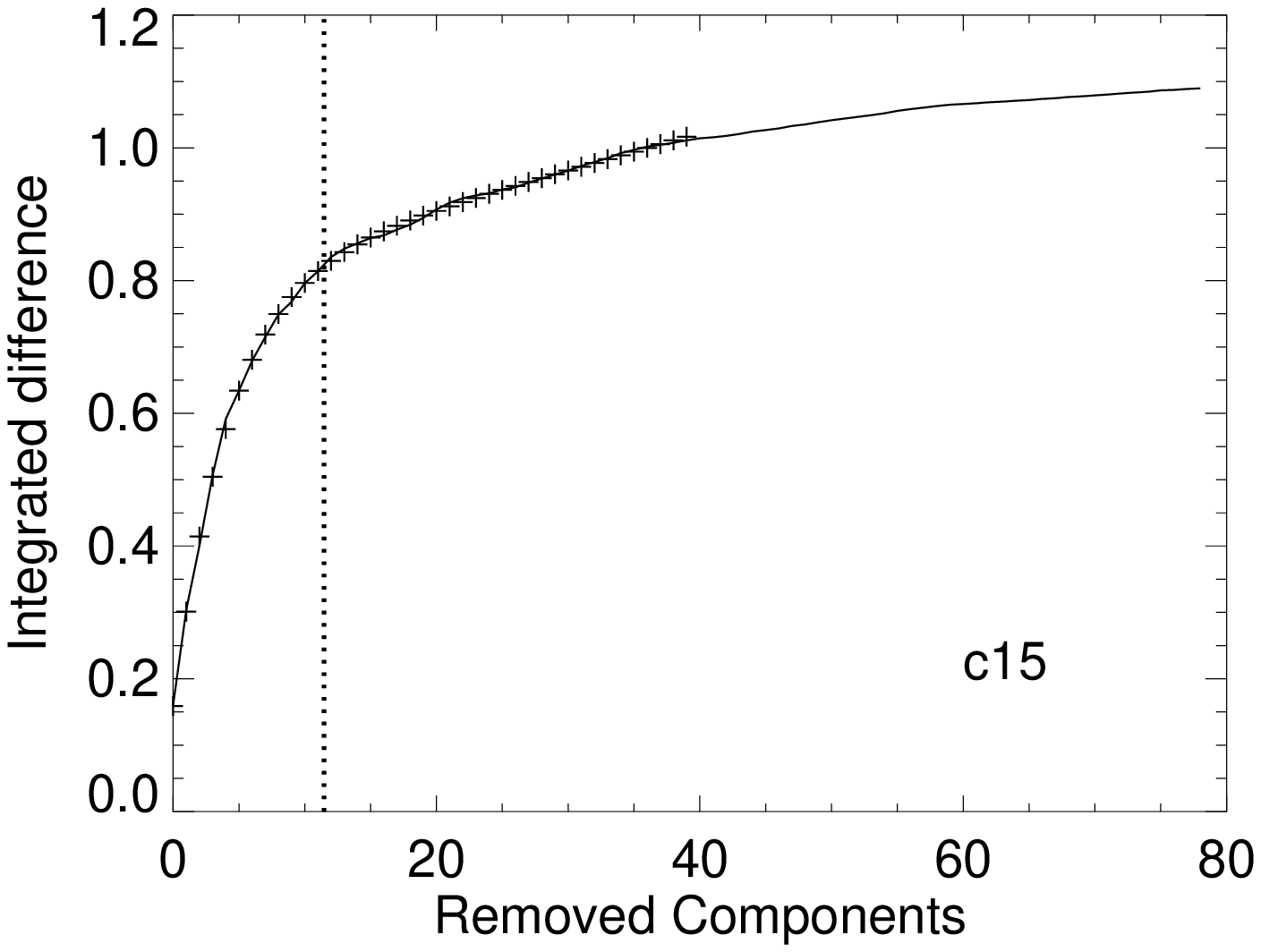}
	\includegraphics[width=0.39\columnwidth ]{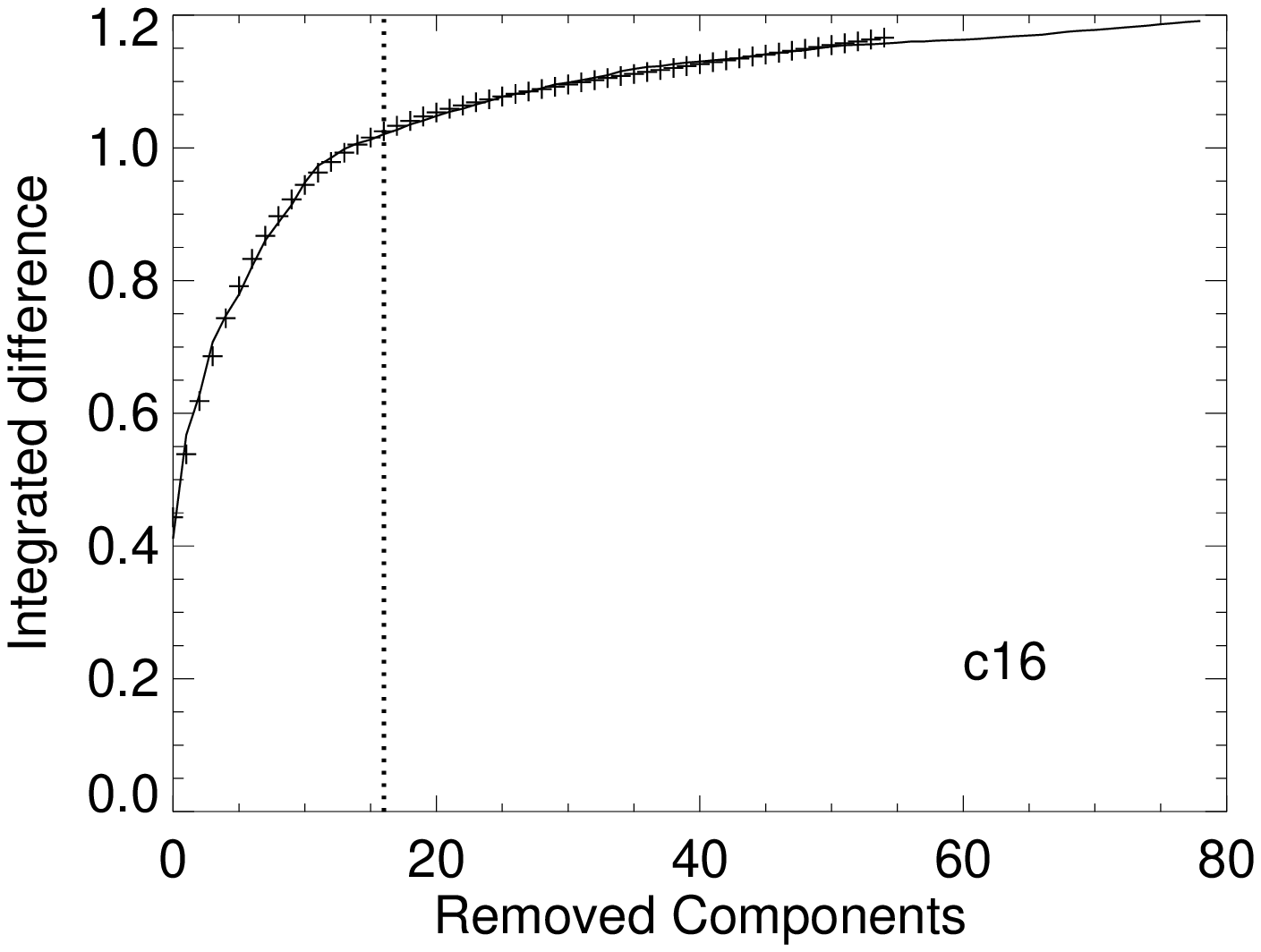}  
	}
	\resizebox{\textwidth}{!}
	{
	\includegraphics[width=0.39\columnwidth ]{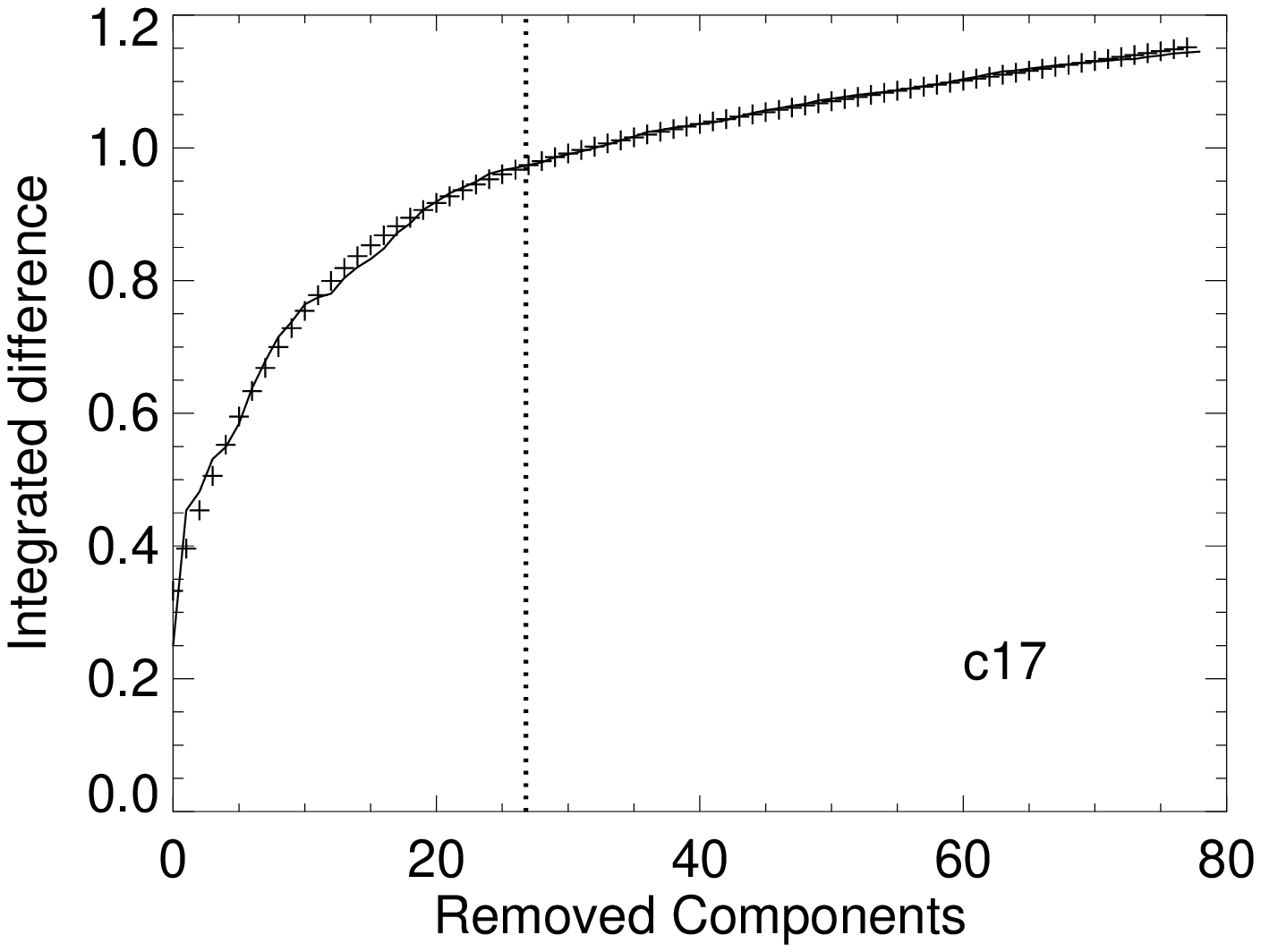}
	\includegraphics[width=0.39\columnwidth ]{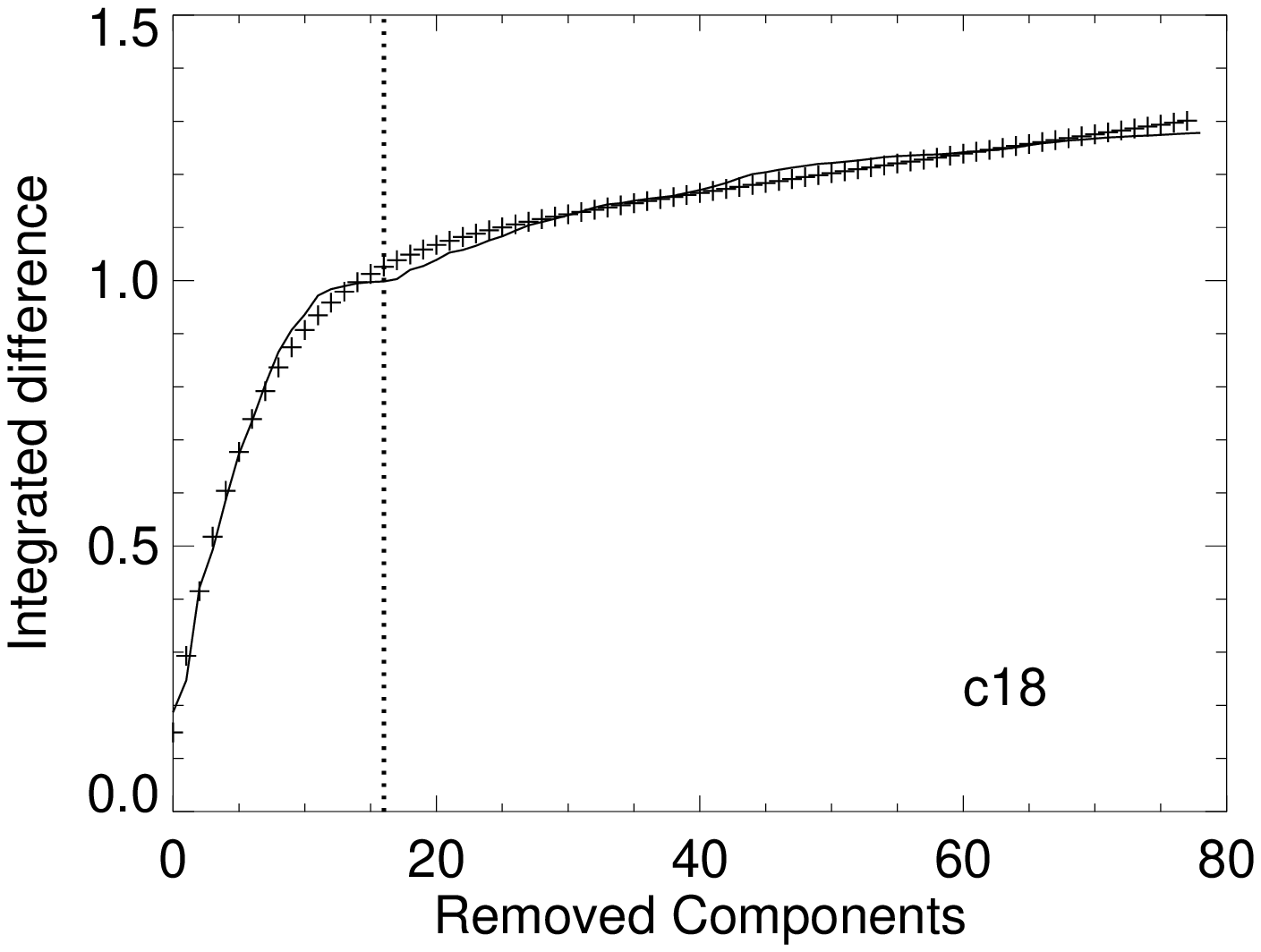}  
	}
	\resizebox{\textwidth}{!}
	{
	\includegraphics[width=0.39\columnwidth ]{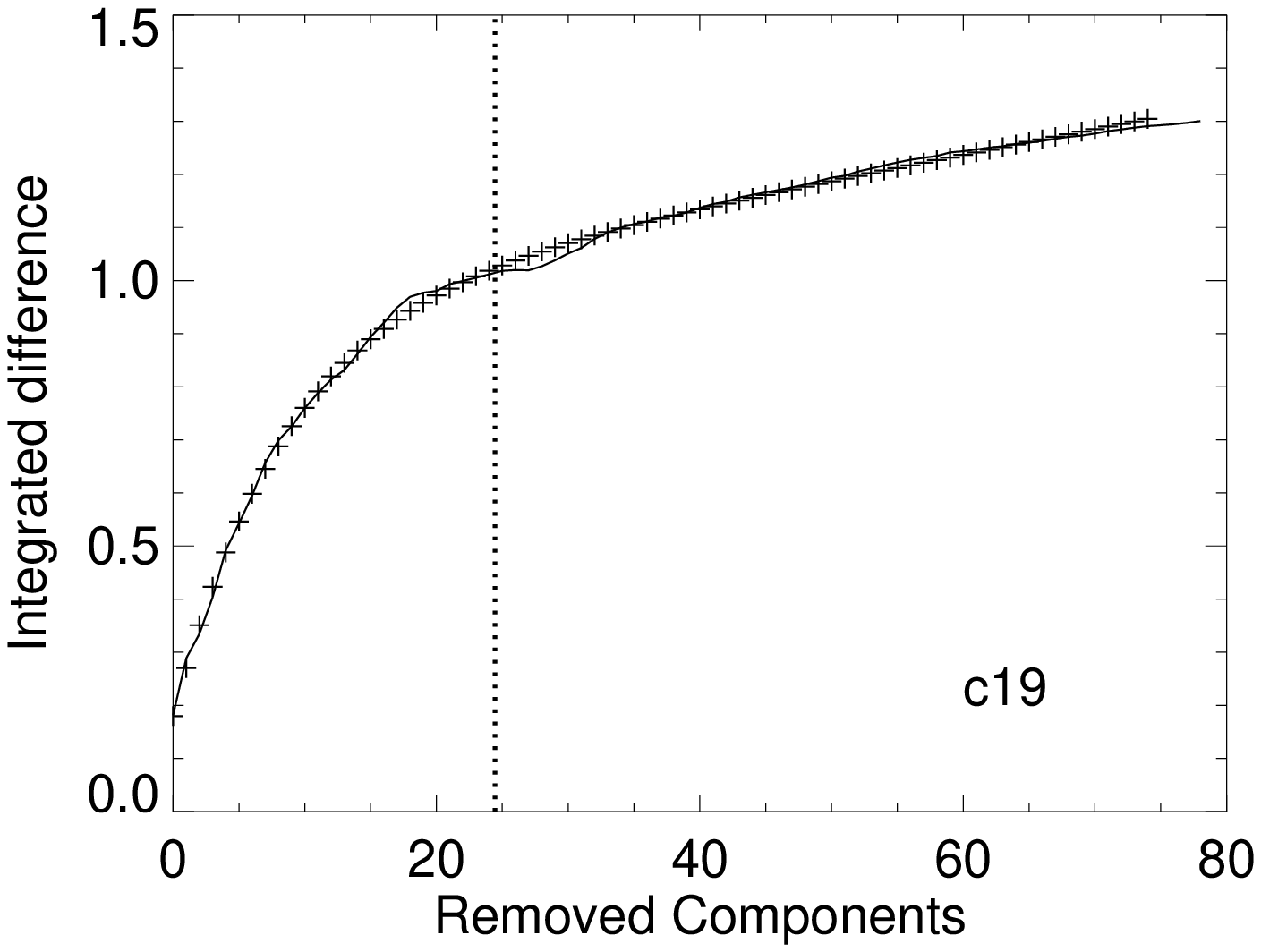}
	\includegraphics[width=0.39\columnwidth]{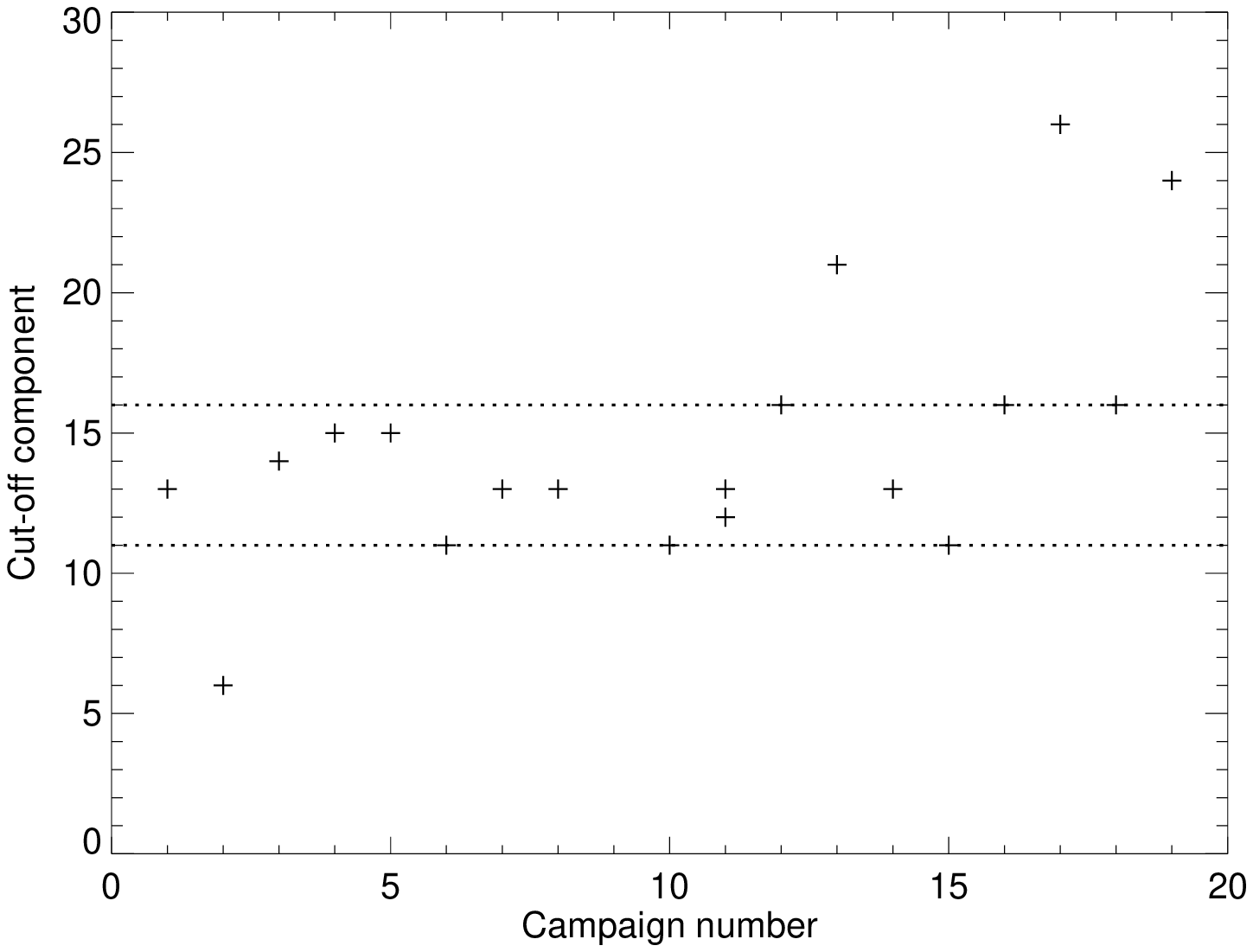}
	}
\caption{{Same as Fig.~\ref{fig:comps} but for Campaign 11b to 19. We also show the cut-off component as a function of the Campaign (bottom right) and two horizontal dashed lines mark the values $11$ and $16$.}}
\label{fig:comps1}
	\end{figure*}

\section{Results}
\label{sec:res}
Once the principal components are computed and the cut-off has been estimated, we are able to remove their effects from the light curves making use of the equation~\ref{eq:recx}. How can we be sure that we remove only/all the systematic effects? Since there is no an unequivocal way to check on the goodness of the light curves, in the following we present different checks that we made in order to validate the method and the reconstructed light curves.

\subsection{Global trends (large temporal scale effects)}

First of all, we consider that, because the systematic effects strongly affect all the original light curves, if we add them all, {we should see large temporal scale effects, resulting in residual global trends. To account for the different flux level and variability of each star, we subtract each light curve by its average flux and divided it by its standard deviation.} Therefore, we compared the summed original light curves with the ones obtained applying our method and, for comparison purpose, with the ones corrected by the built-in procedure, the PDC, {and their standard deviations}. The result of this comparison is presented in Fig.~\ref{fig:globthr} and Fig.~\ref{fig:globthr1}.
 As expected, a global trends is clearly visible in the original light curve in each campaign. A small residual trend is also visible on the PDC corrected light curve, meaning that although the systematic affects are strongly mitigated, they are still present. However, if we look at the curves obtained applying the PCA method, they appear, generally, flat with small oscillations, meaning that light curves after the correction are reasonably uncorrelated. This does not apply to Campaign 2 for which systematics are still present and, indeed, PCA selects a very small number of components to be removed ($6$). We also report a strange behaviour in Campaign 19 in the case for the PDC correction that shows a trend much pronounced than the original light curves. {If we exclude Campaign 2 and 19 for which PDC and PCA show anomalous behaviours, on average the PCA reduces the standard deviation of the light curves with respect the PDC by $\sim40$\% and the scatter of the global trends by a factor of $\sim 3$ (see Fig.~\ref{fig:globthr1} bottom-right).}

\begin{figure*}[!th]
	\centering
	\resizebox{\textwidth}{!}
	{
	\includegraphics[width=0.39\columnwidth ]{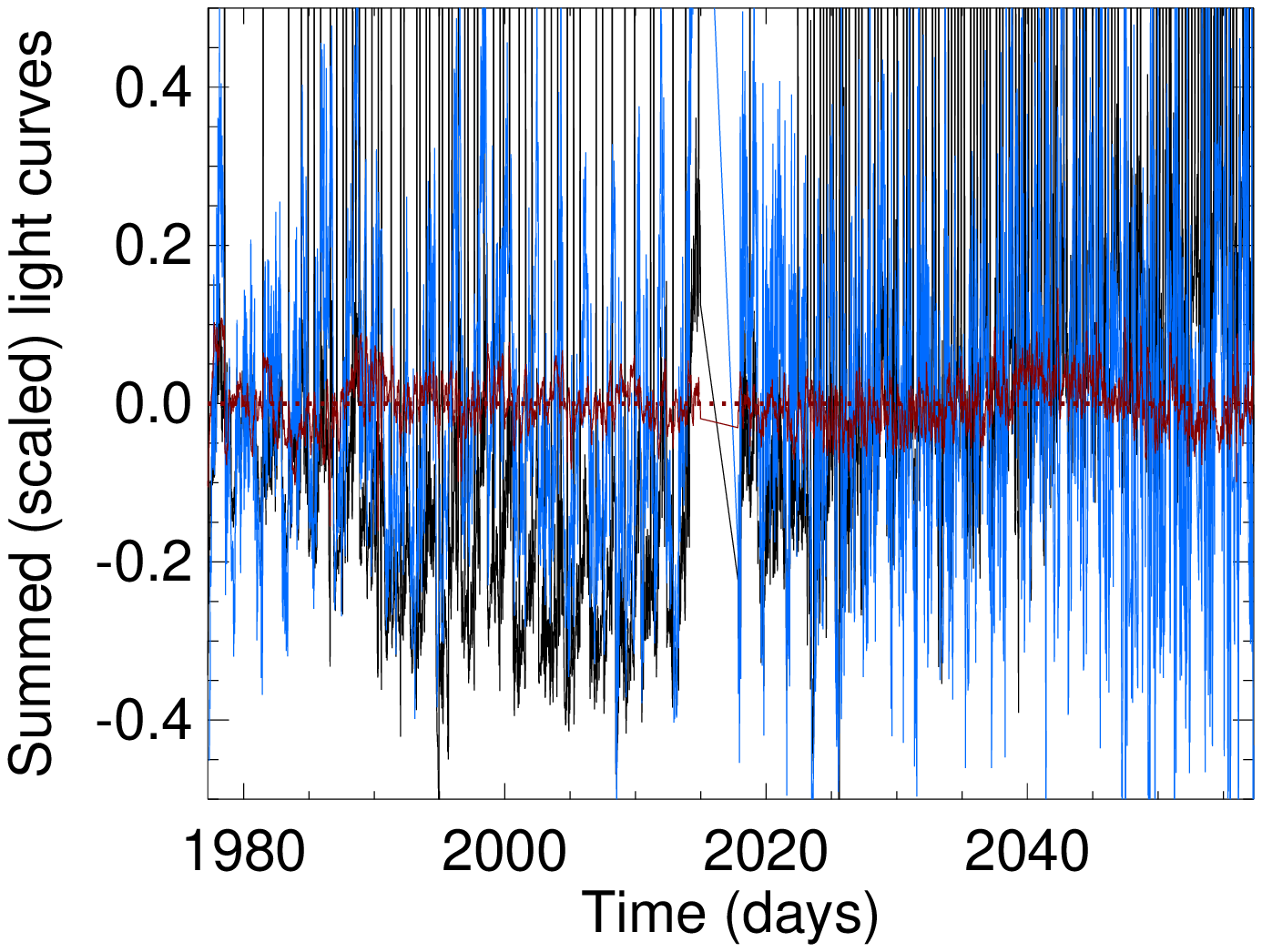}  
	\includegraphics[width=0.39\columnwidth ]{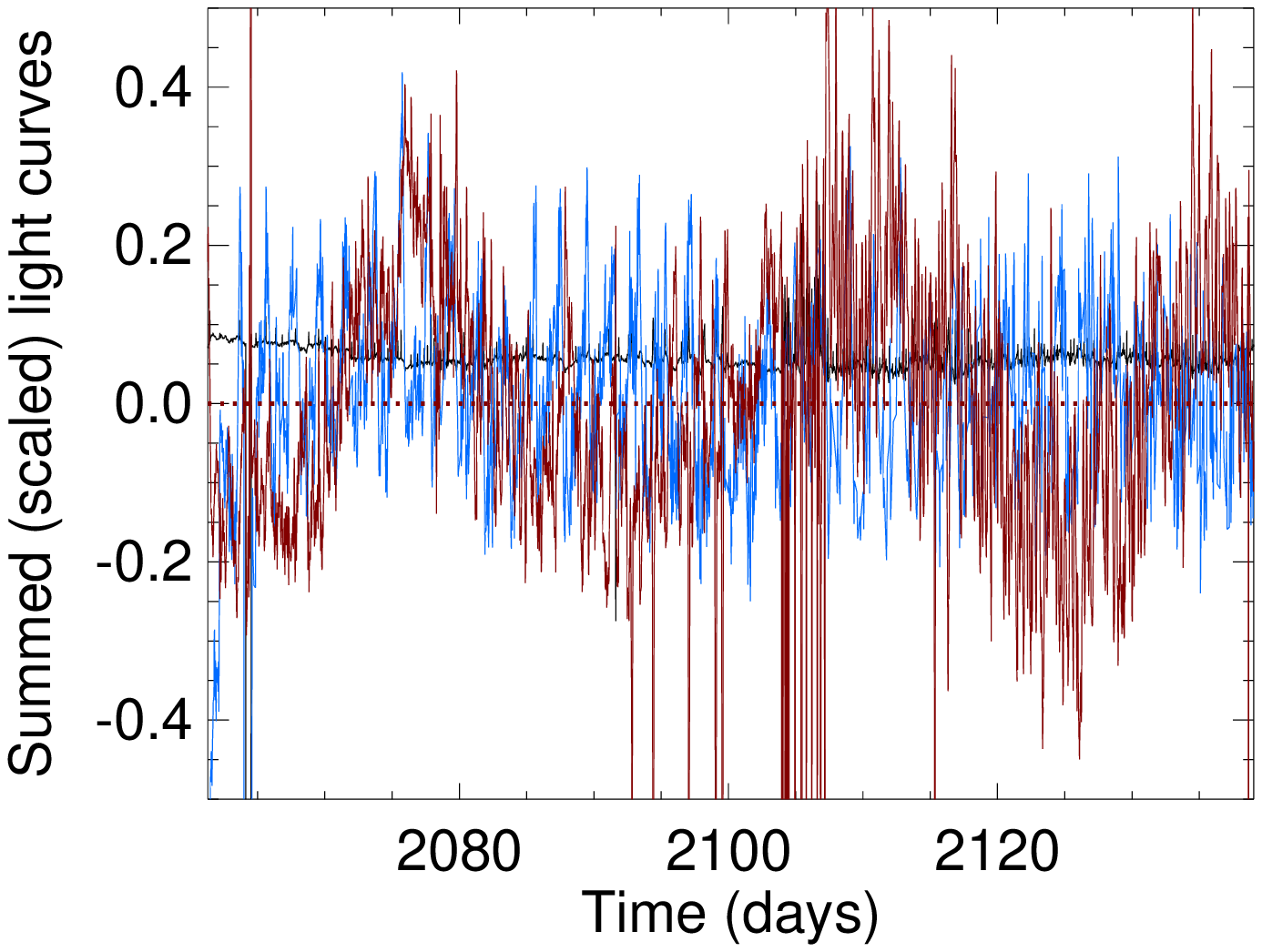}
	}
	\resizebox{\textwidth}{!}
	{
	\includegraphics[width=0.39\columnwidth ]{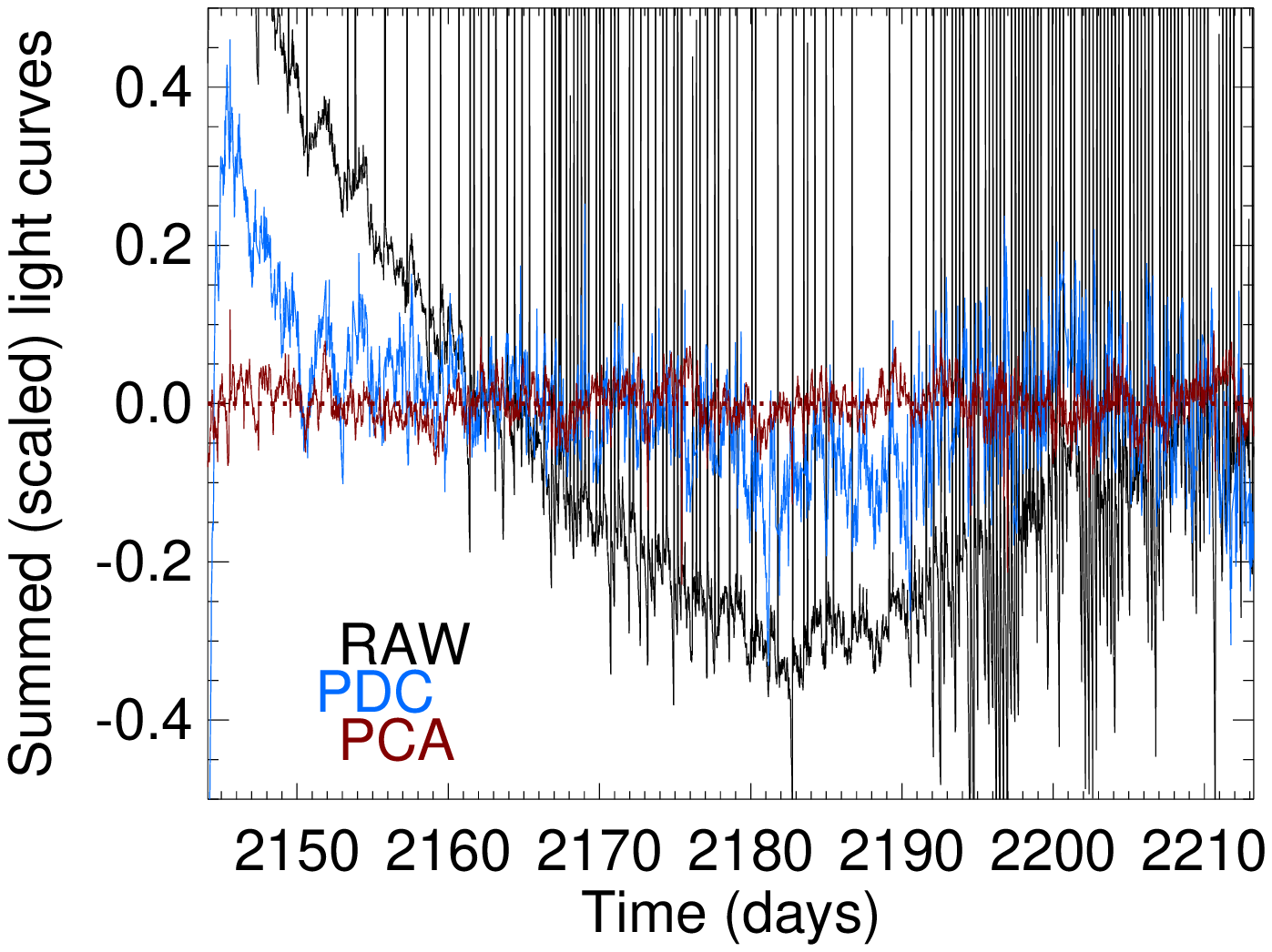}
	\includegraphics[width=0.39\columnwidth ]{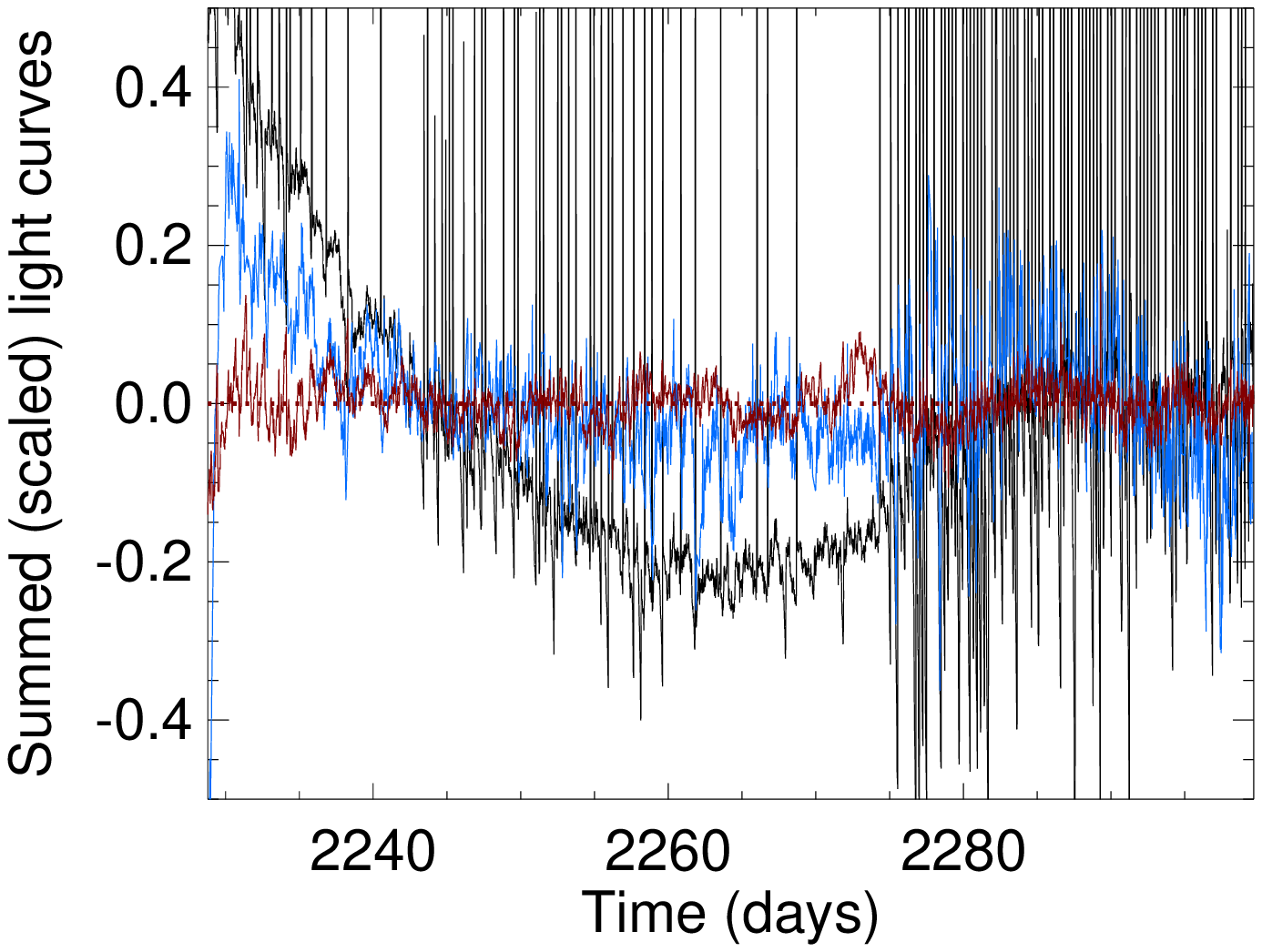}  
	}
	\resizebox{\textwidth}{!}
	{
	\includegraphics[width=0.39\columnwidth ]{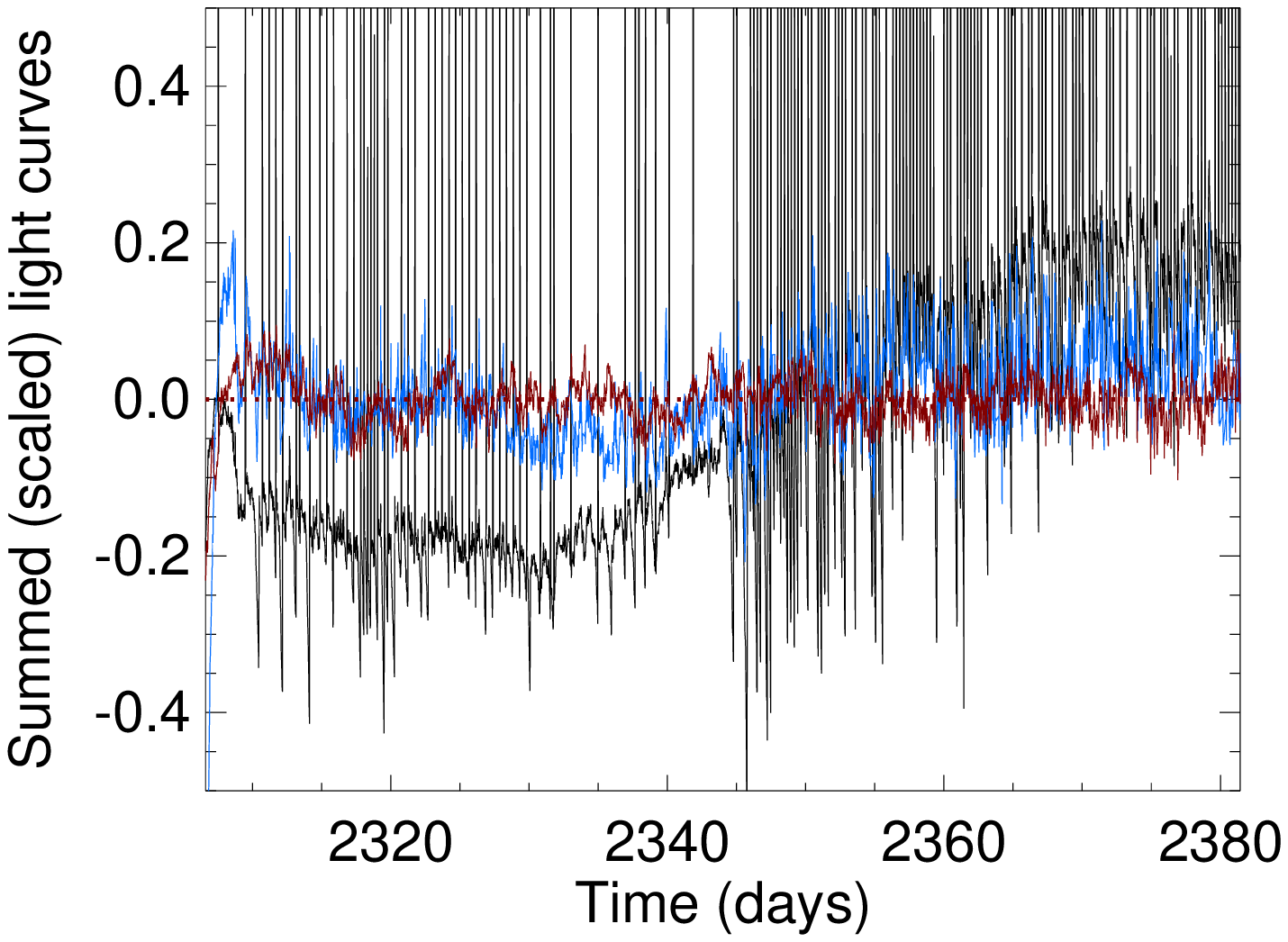}
	\includegraphics[width=0.39\columnwidth ]{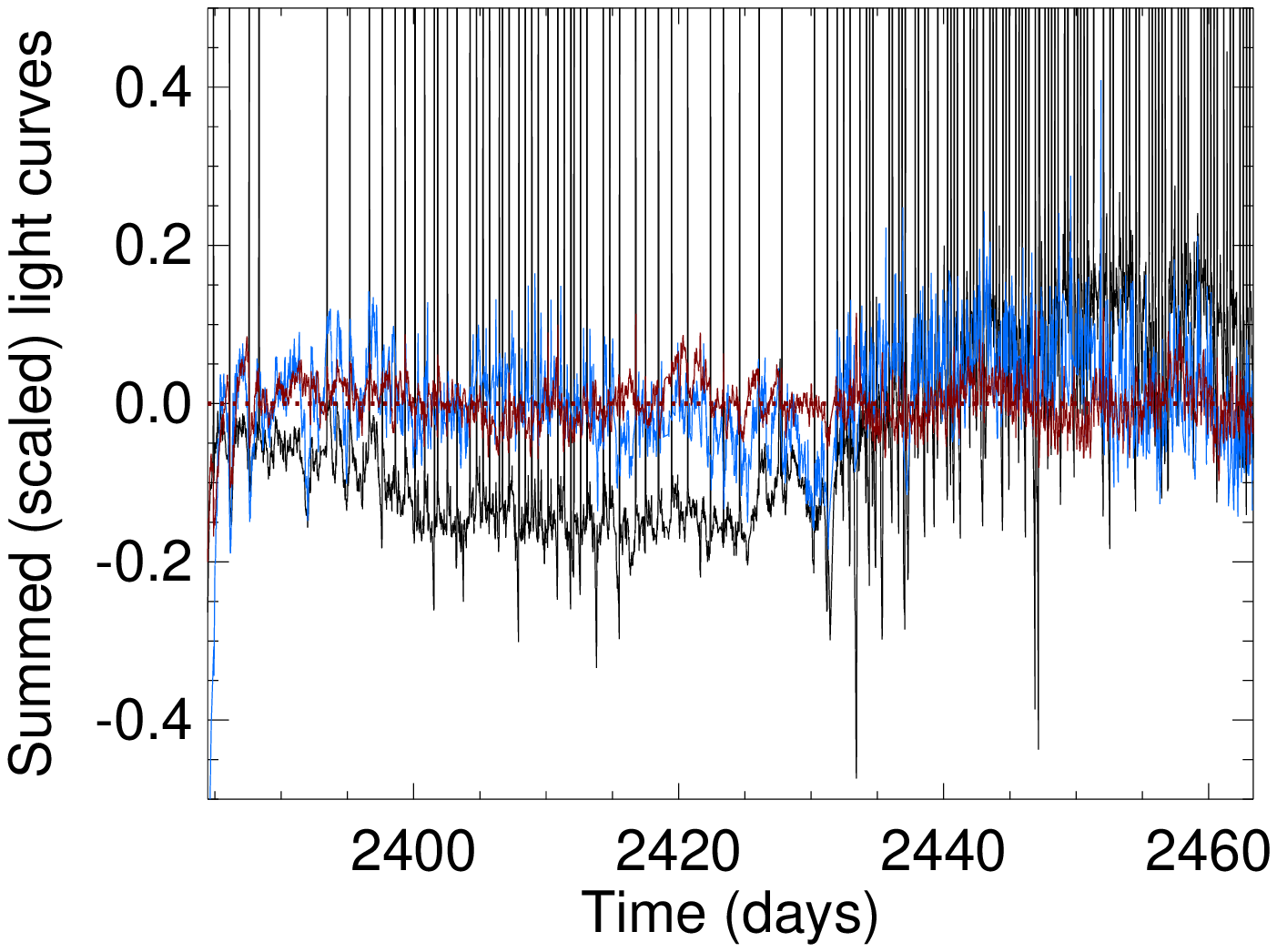}  
	}
	\resizebox{\textwidth}{!}
	{
	\includegraphics[width=0.39\columnwidth ]{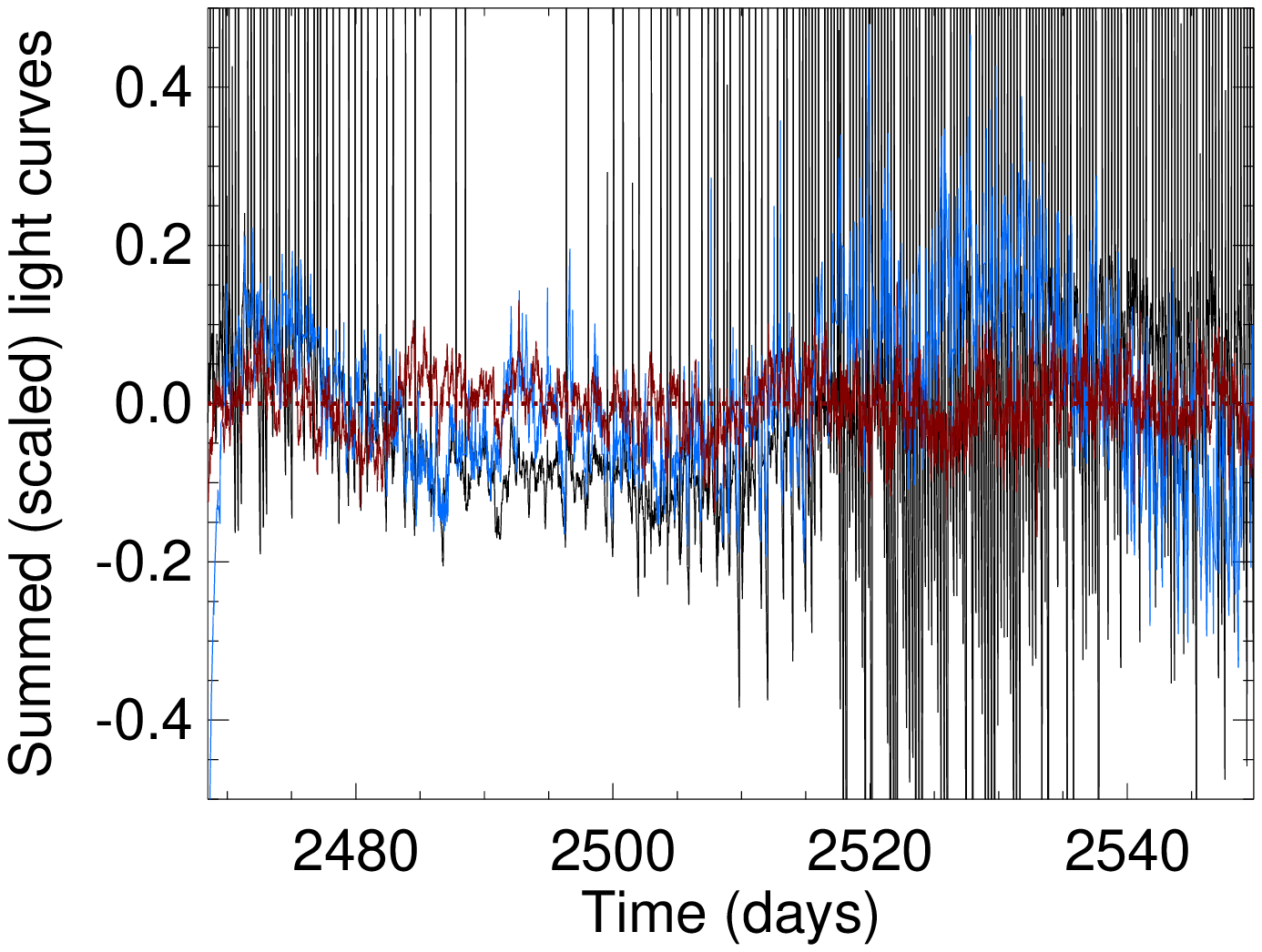}
	\includegraphics[width=0.39\columnwidth ]{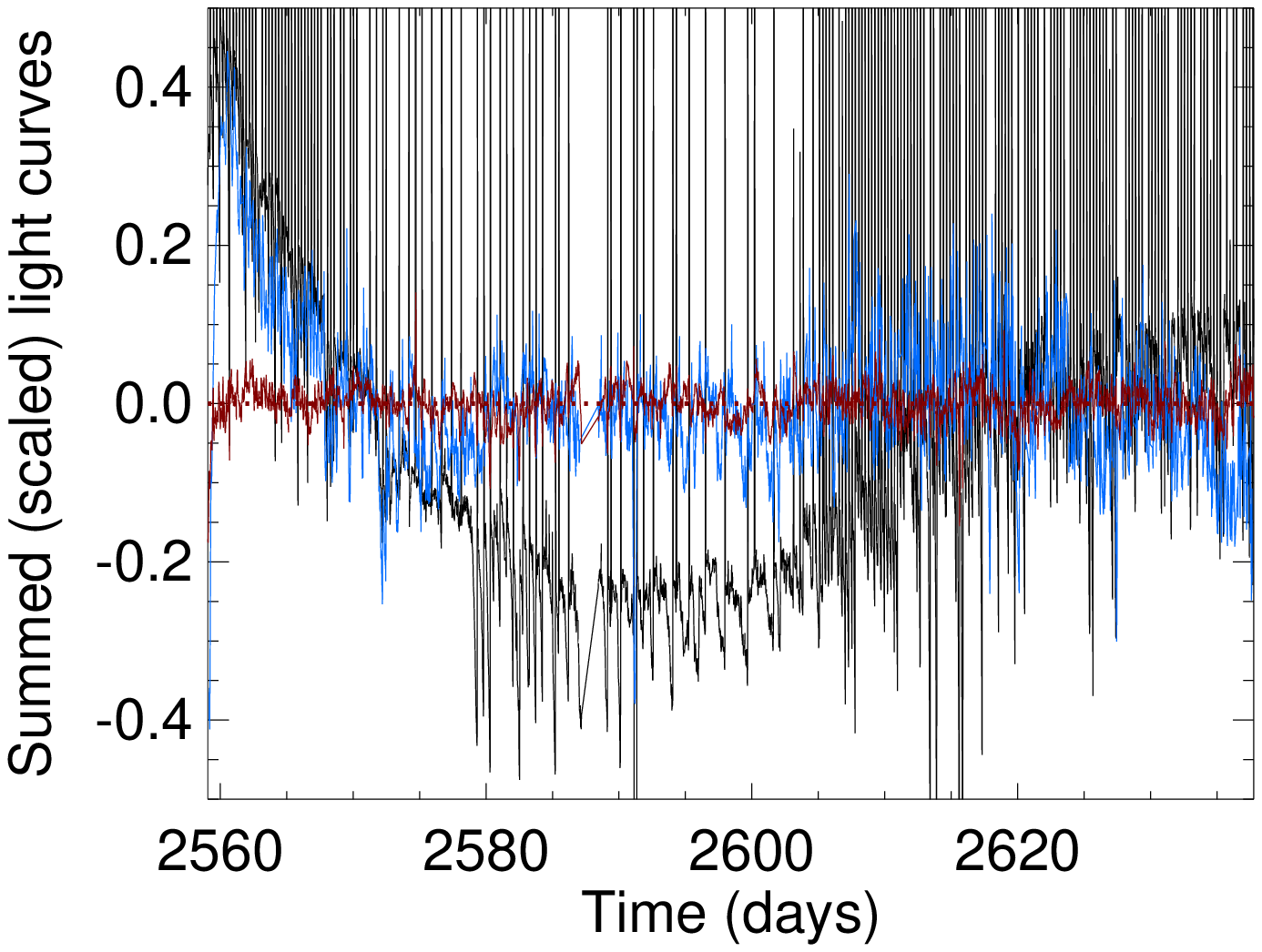}  
	}
	\resizebox{\textwidth}{!}
	{
	\includegraphics[width=0.39\columnwidth ]{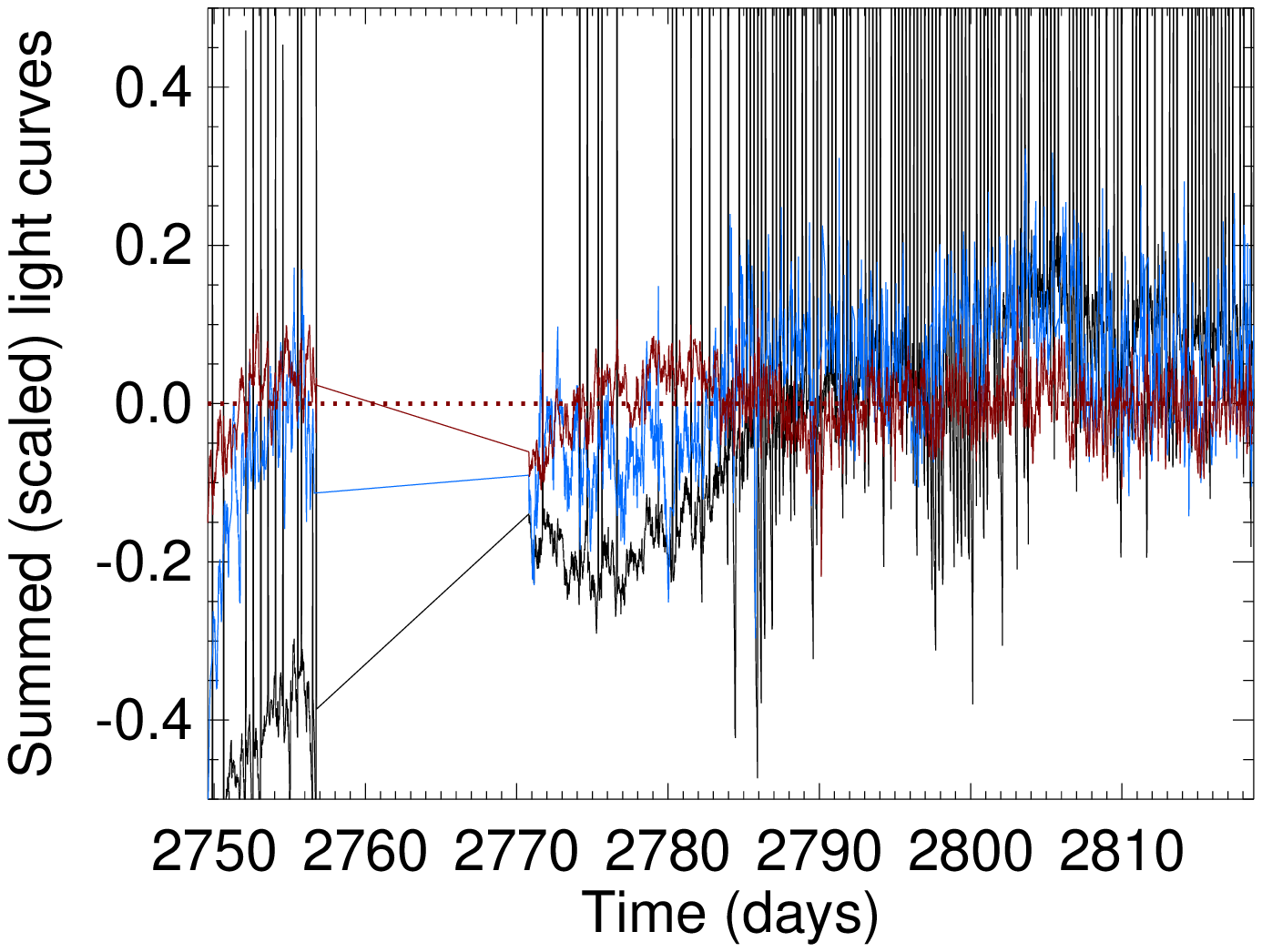}
	\includegraphics[width=0.39\columnwidth ]{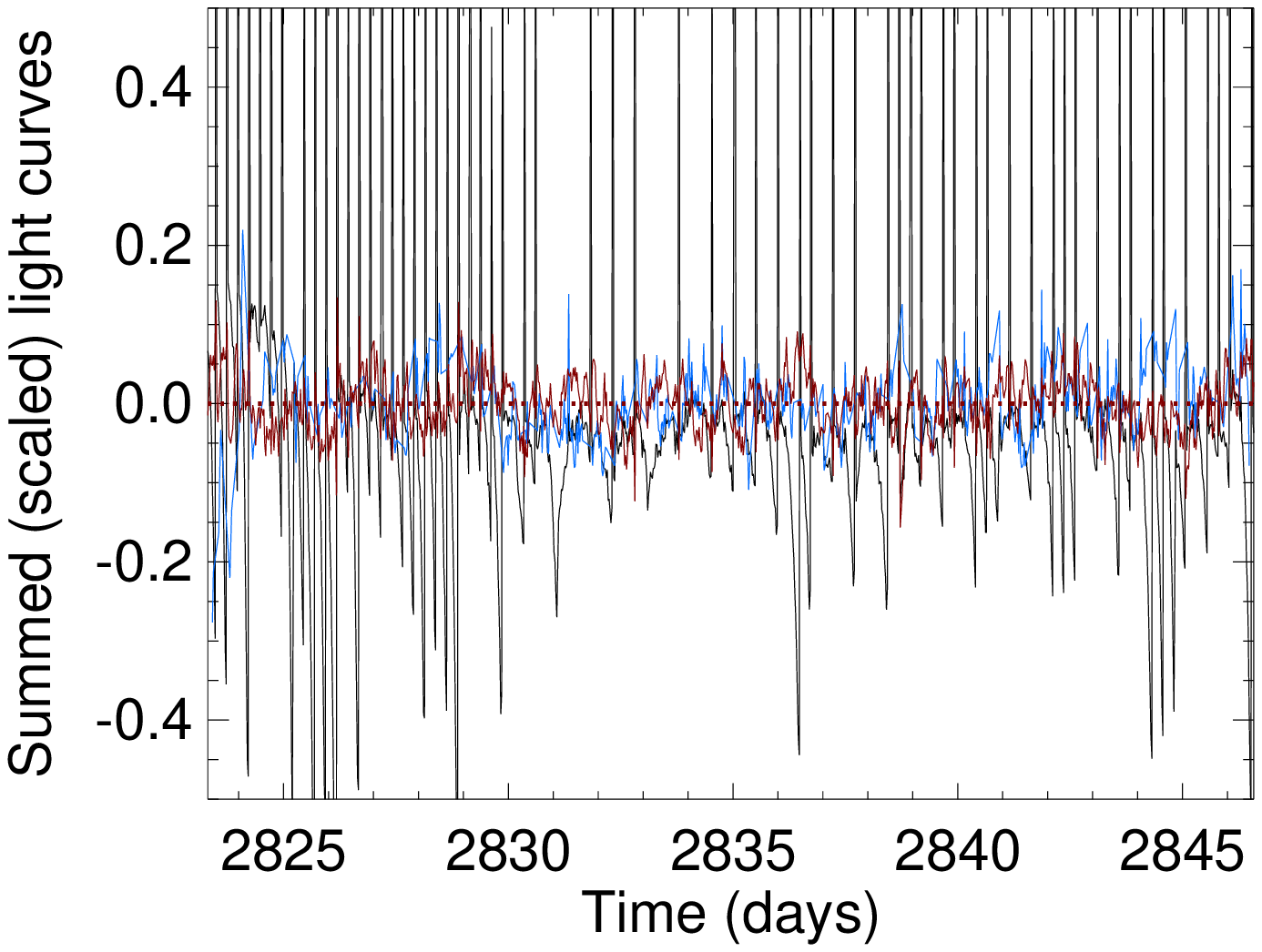}  
	}
\caption{Global trends obtained by adding all the light curves (shifted by their mean and divided by their standard deviation) of a campaign for the original data (black line), PDC corrected (blue dashed line) and PCA corrected data (red line), for all the campaign. A straight line (dashed red line) $y=0$ is also shown for clearness {and the figures order is the same as in Fig.~\ref{fig:comps}}.}
\label{fig:globthr}
	\end{figure*}

\begin{figure*}[!th]
	\centering
	\resizebox{\textwidth}{!}
	{
	\includegraphics[width=0.39\columnwidth ]{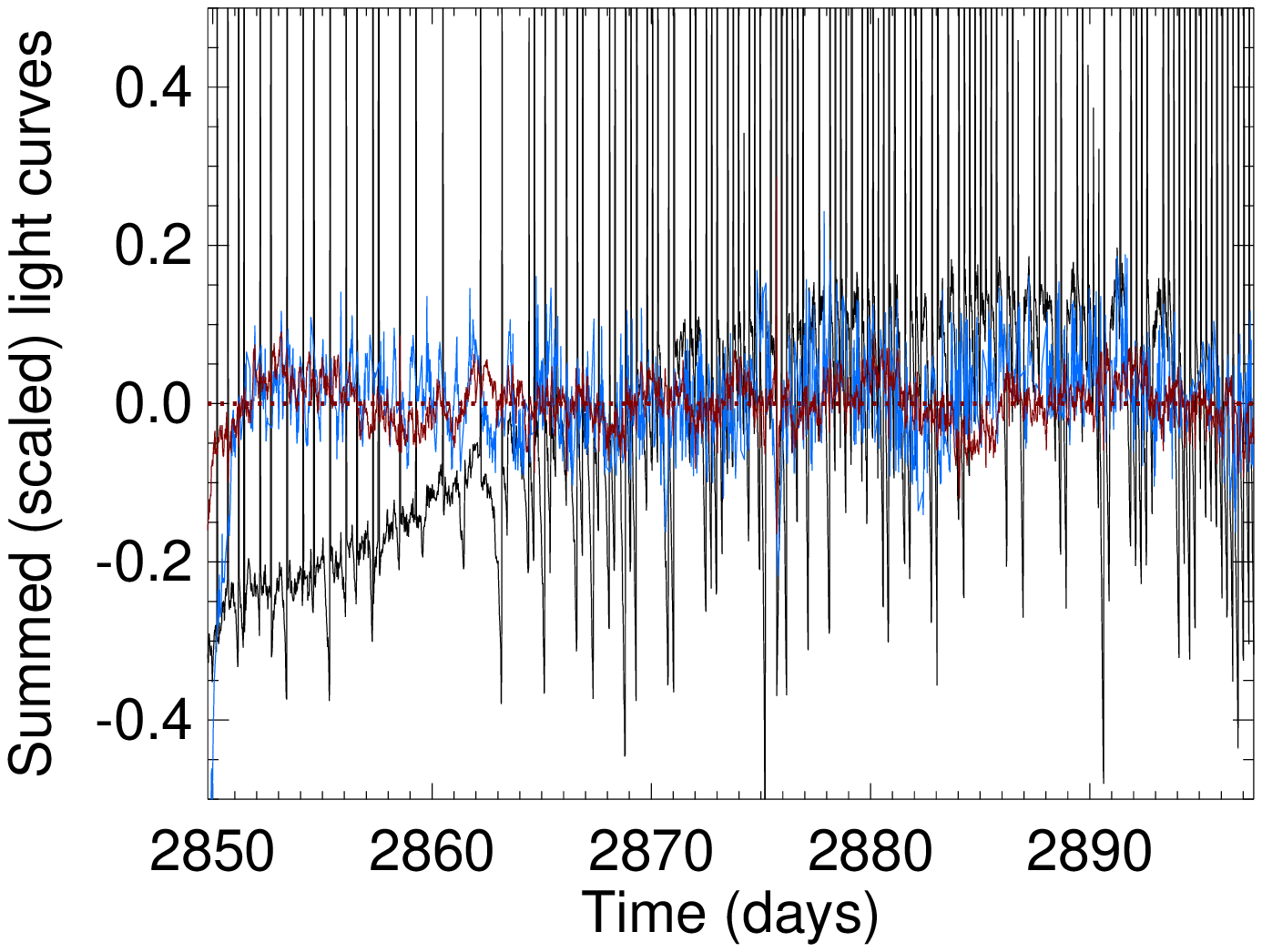}  
	\includegraphics[width=0.39\columnwidth ]{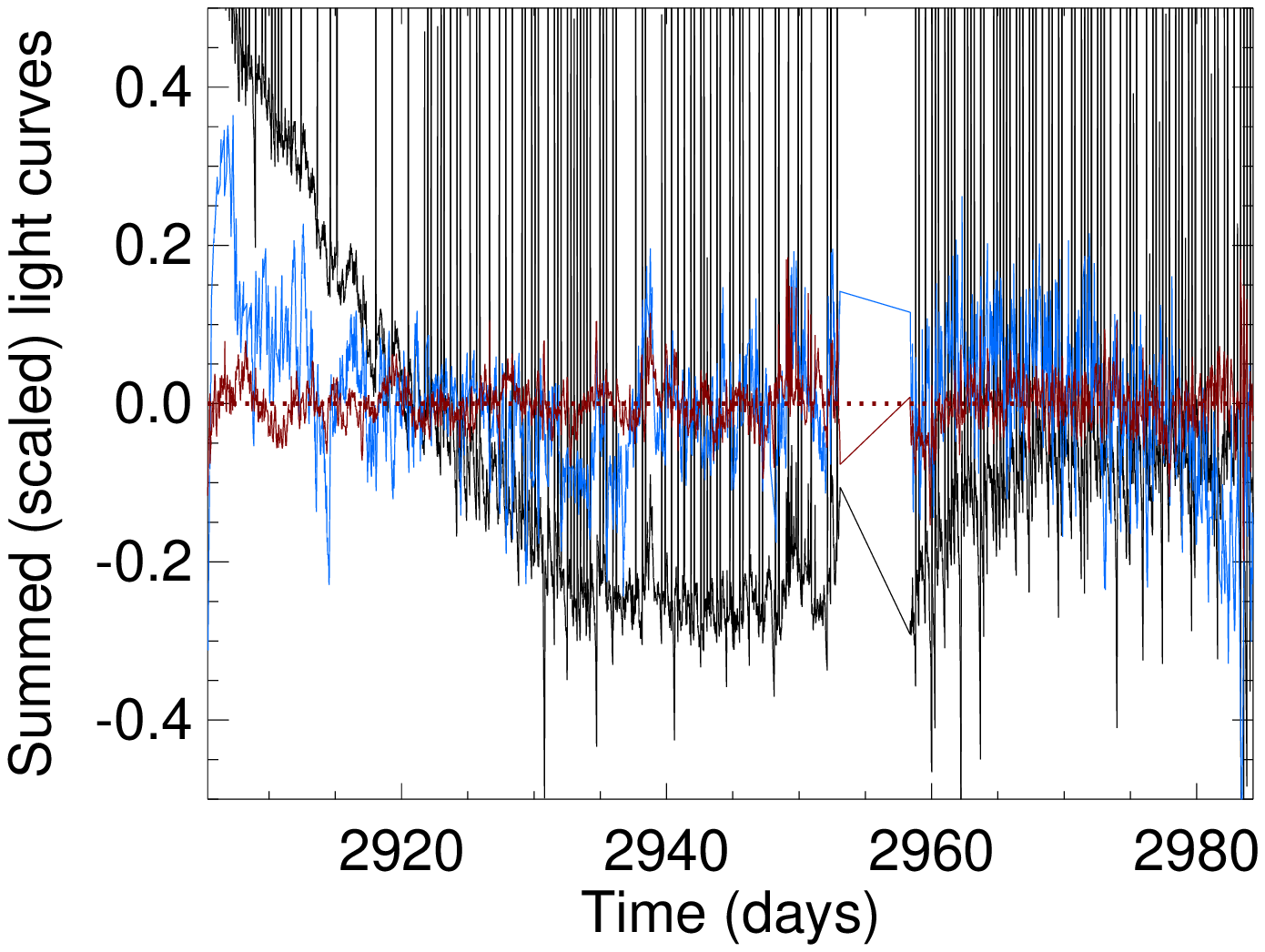}
	}
	\resizebox{\textwidth}{!}
	{
	\includegraphics[width=0.39\columnwidth ]{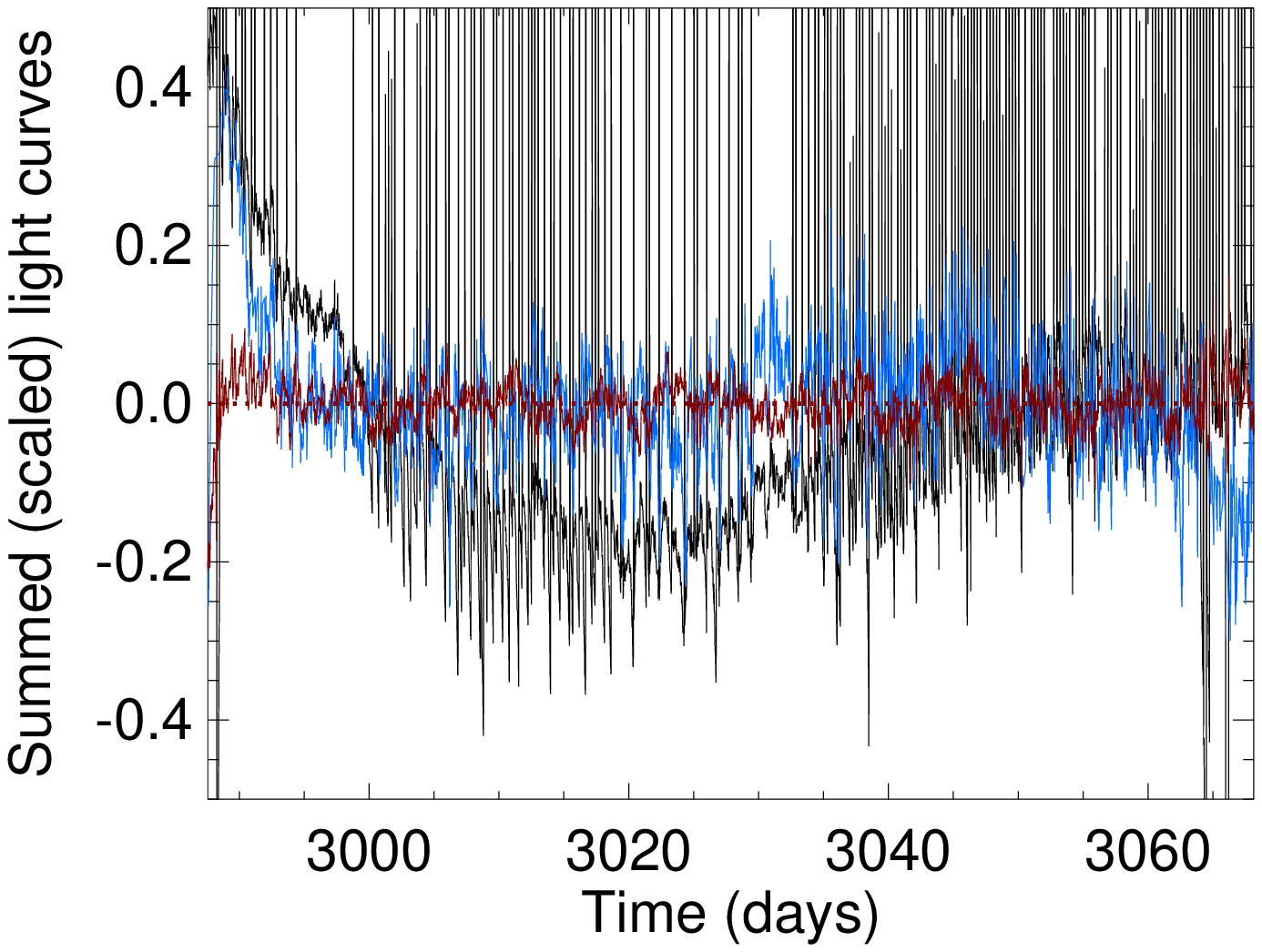}
	\includegraphics[width=0.39\columnwidth ]{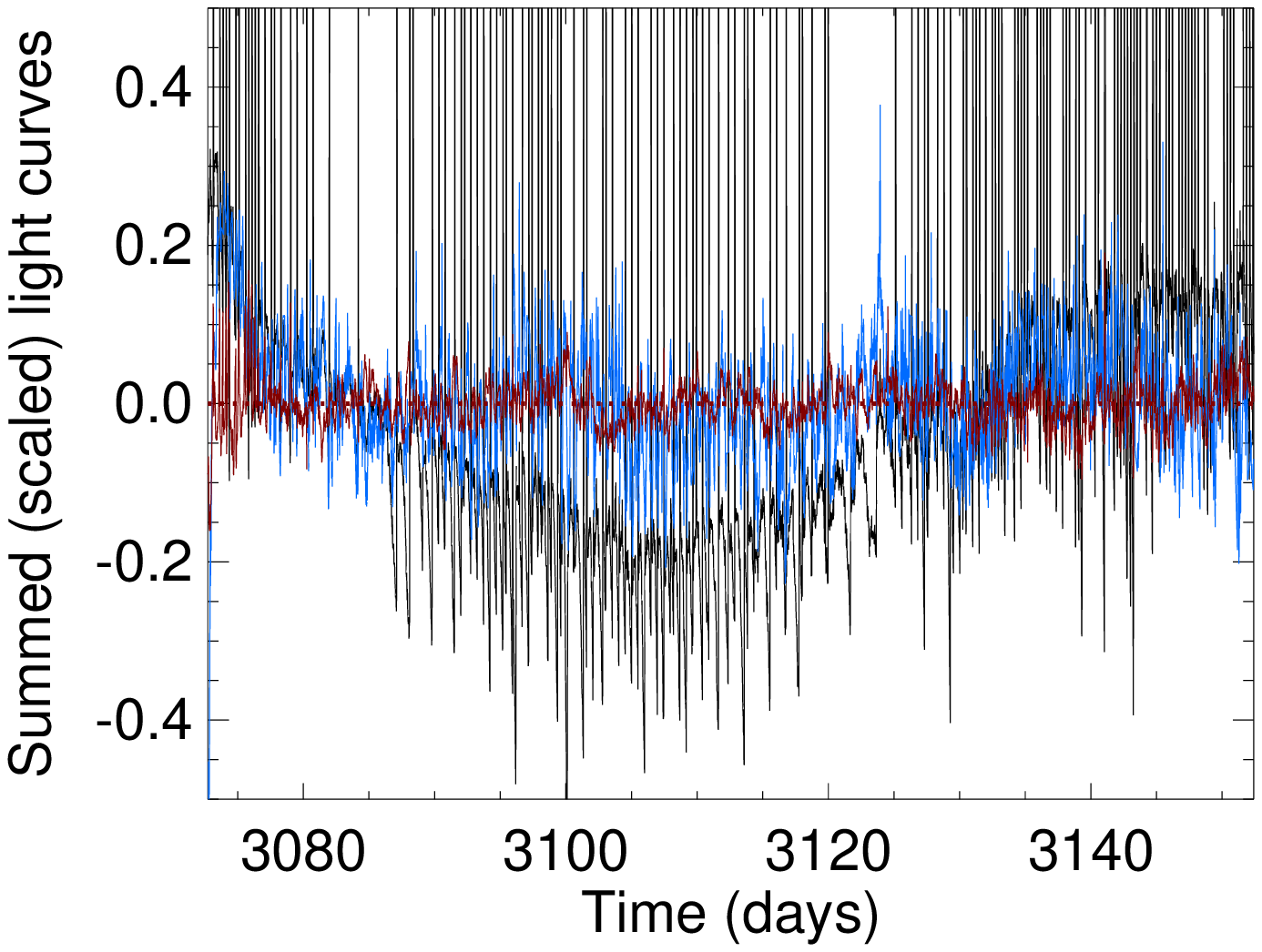}  
	}
	\resizebox{\textwidth}{!}
	{
	\includegraphics[width=0.39\columnwidth ]{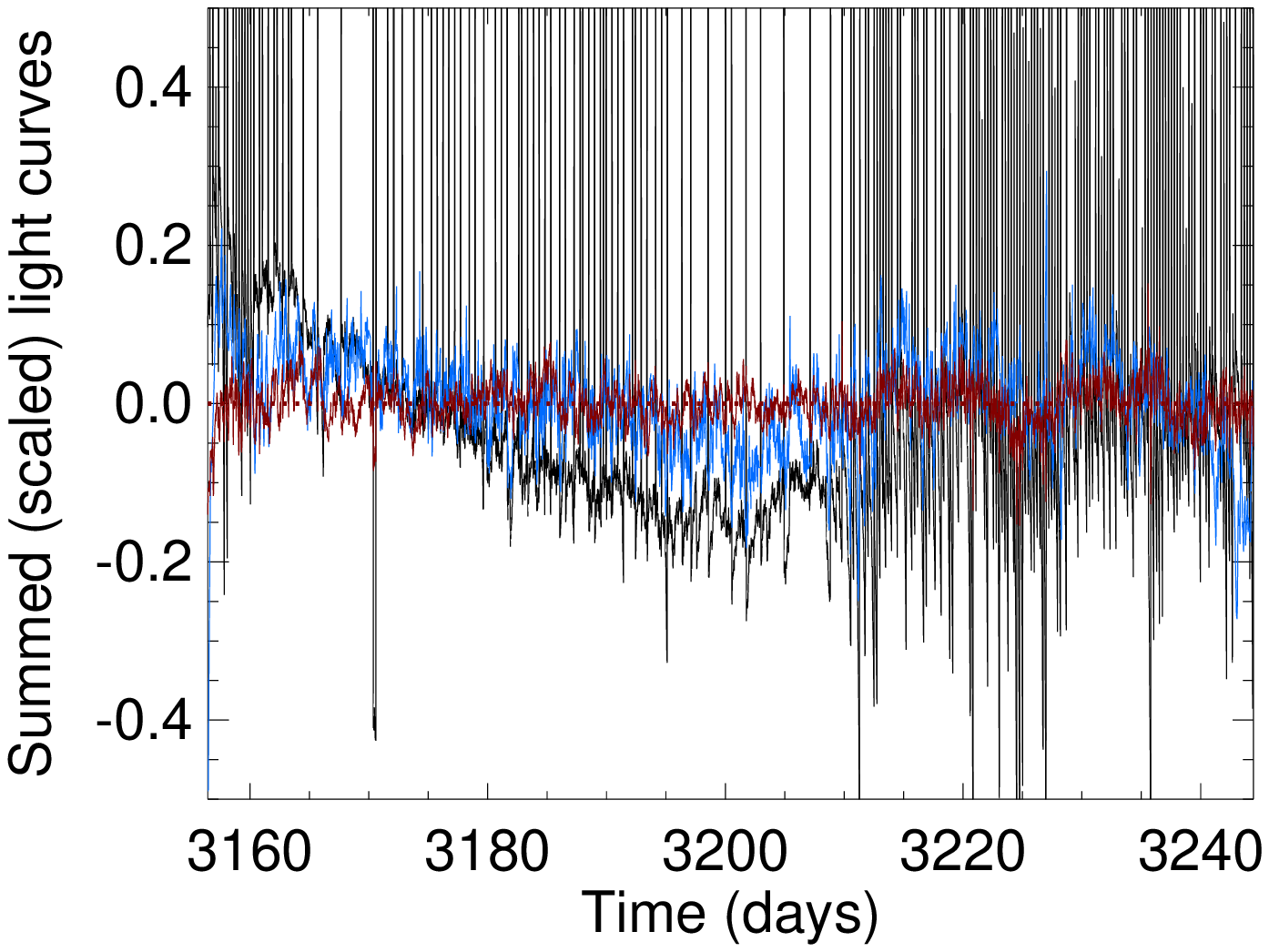}
	\includegraphics[width=0.39\columnwidth ]{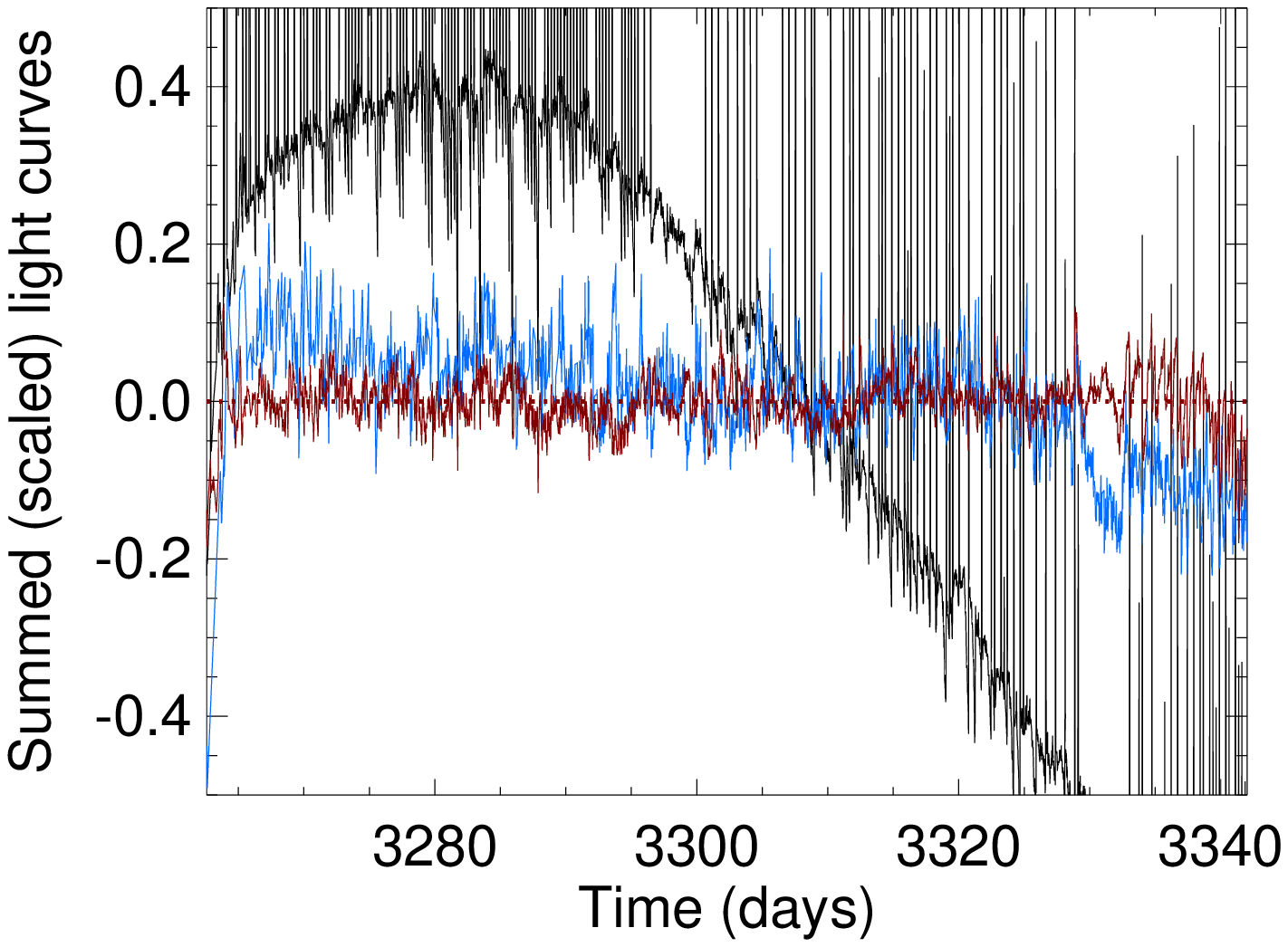}  
	}
	\resizebox{\textwidth}{!}
	{
	\includegraphics[width=0.39\columnwidth ]{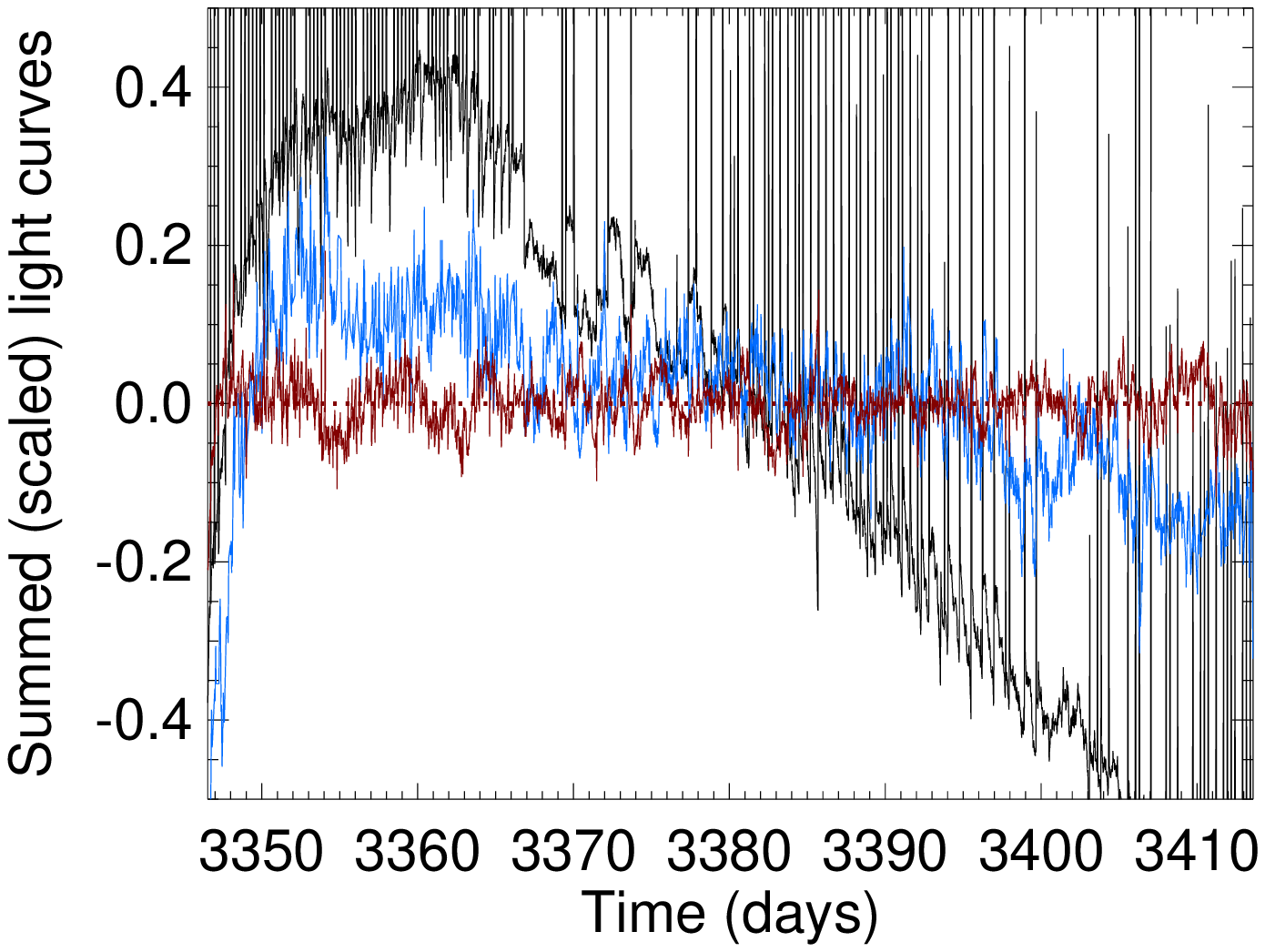}
	\includegraphics[width=0.39\columnwidth ]{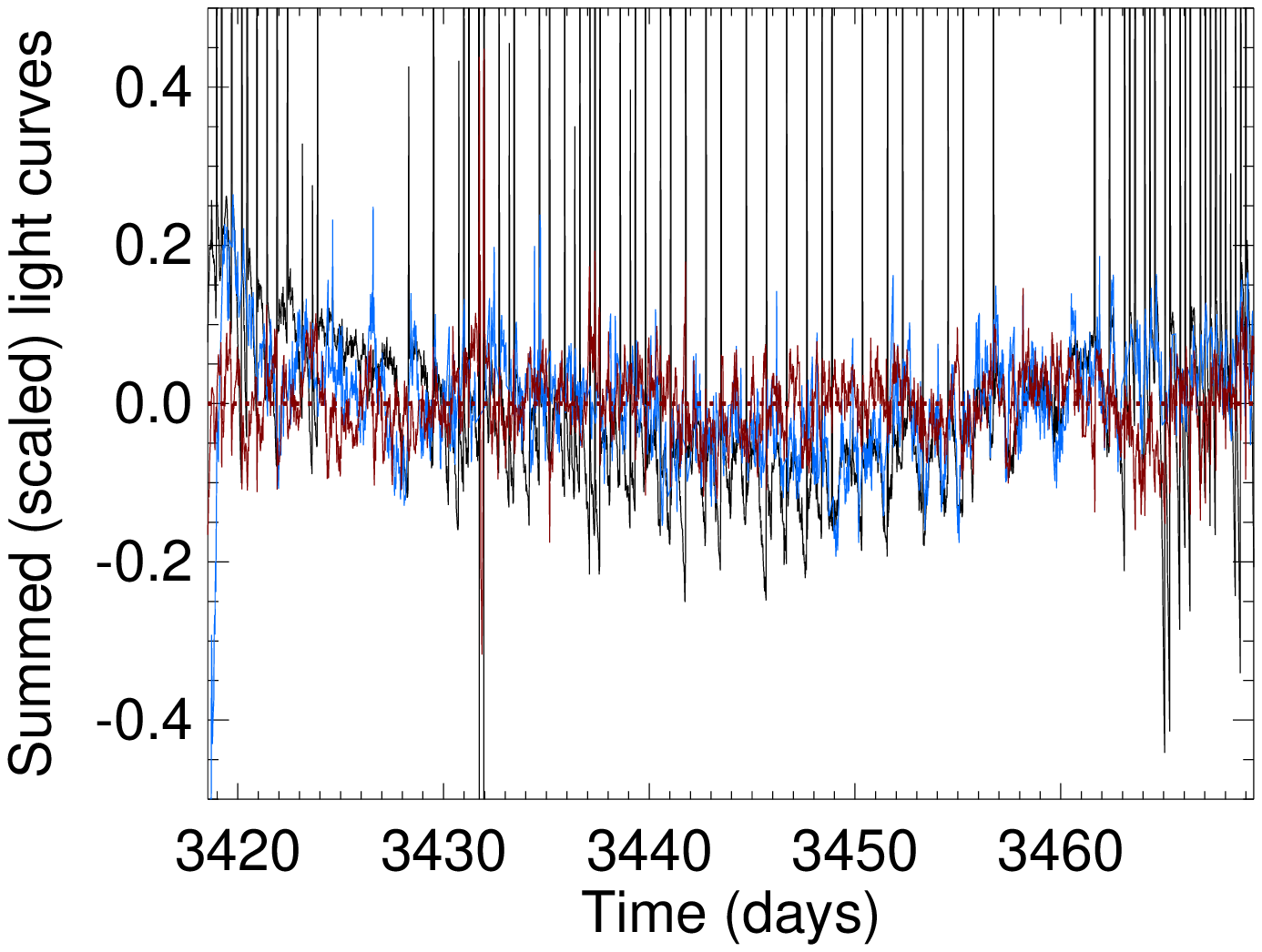}  
	}
	\resizebox{\textwidth}{!}
	{
	\includegraphics[width=0.39\columnwidth ]{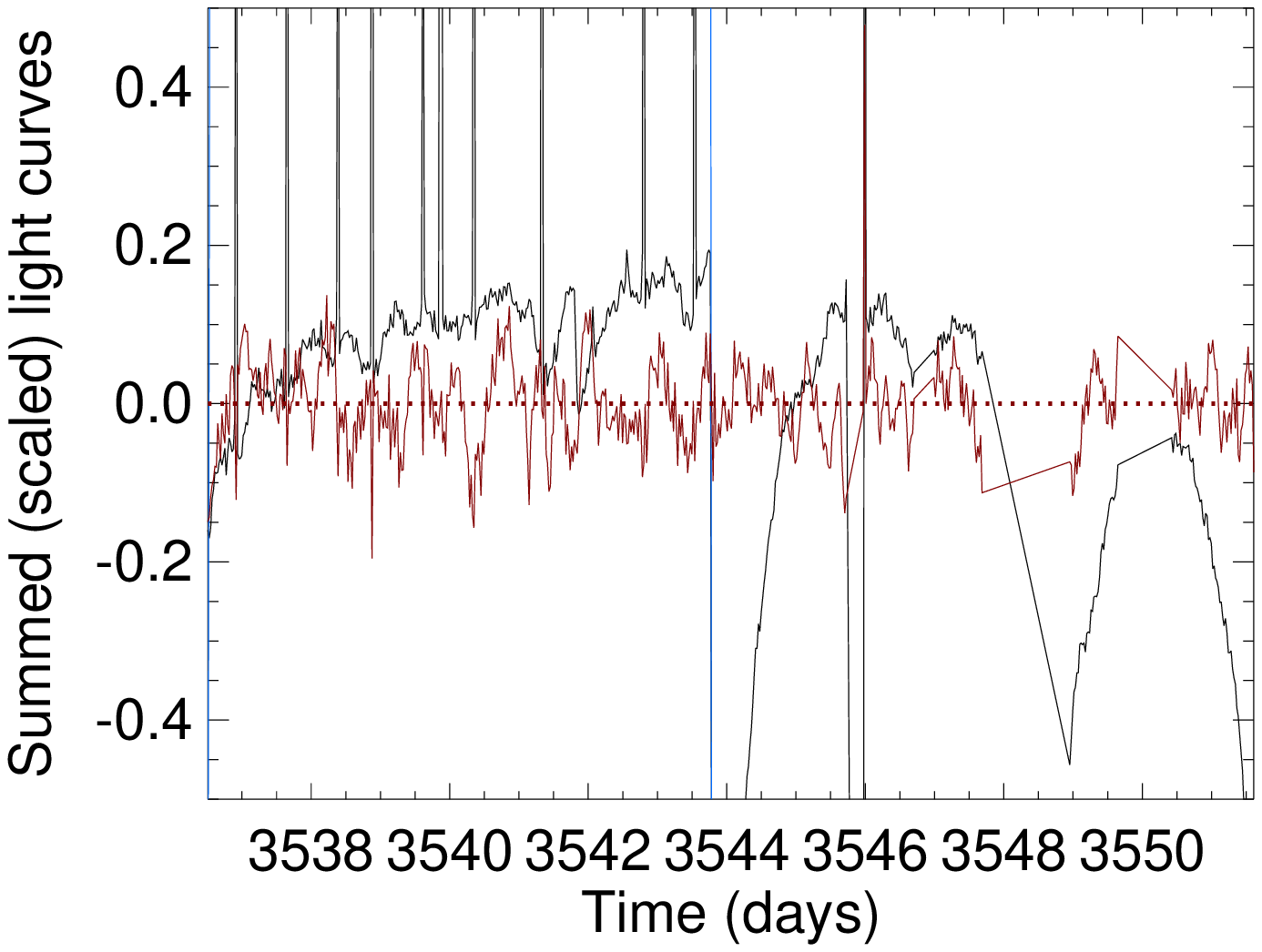}
	\includegraphics[width=0.39\columnwidth ]{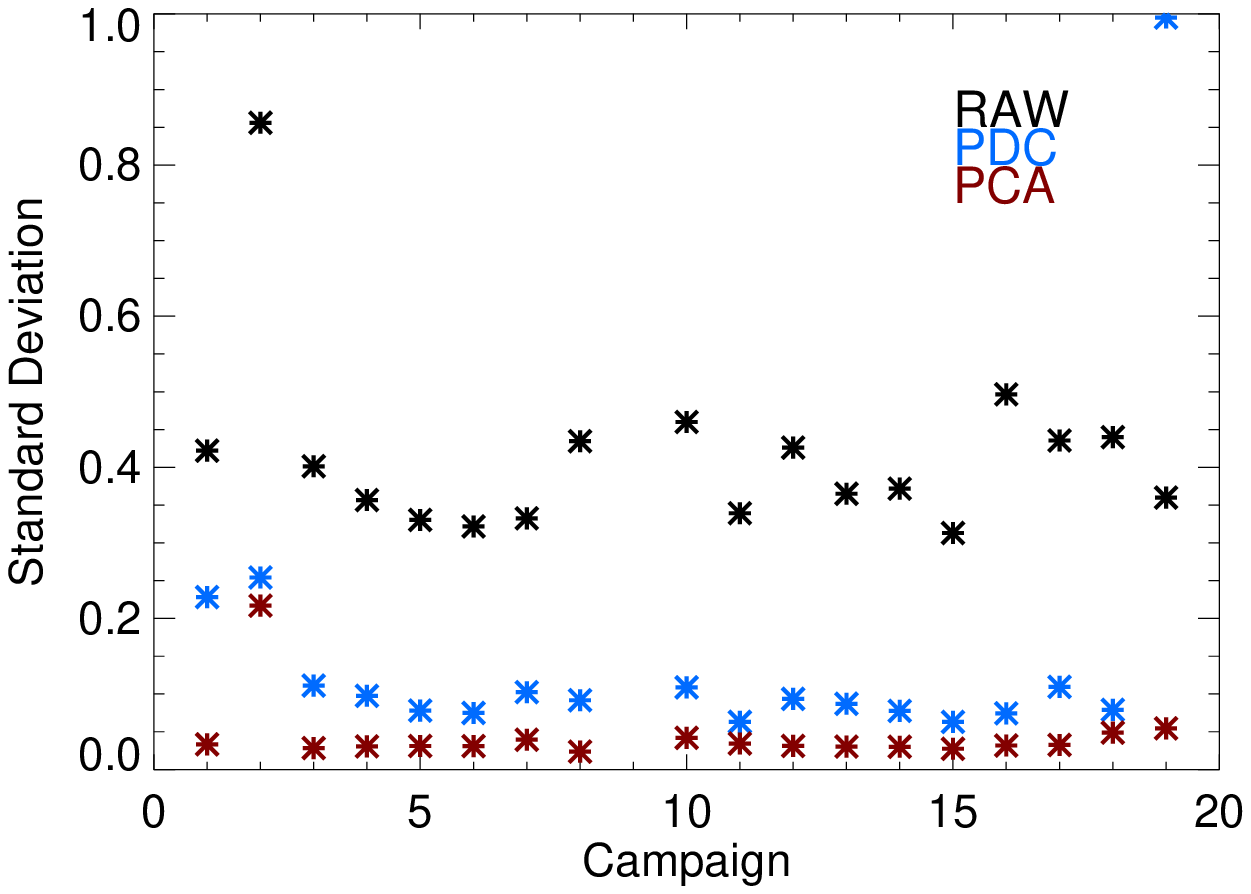}  
	}
\caption{Same as Fig.~\ref{fig:globthr} but for Campaign 11b to 19 {and, in the bottom panel, the standard deviation of the global trends as a function of the campaign. In the case of the Campaign $11$ we average the two values obtained for the campaign 11a and 11b}.}
\label{fig:globthr1}
	\end{figure*}

\subsection{A binary Eclipsing system (EPIC 211135350)}

\label{sec:binecl}

Another check is on the ability of the method in preserving the astrophysical signal after the removal of the systematics. In Fig.~\ref{fig:binecl} we show a light curves of a binary eclipsing system (EPIC $211135350$ in Campaign $4$). We compare the original (raw) data (top-left panel) with the corrected ones (top-right panel) by the PDC and our PCA. As can be seen from the original data, strong spikes occur all over the observational window, stronger in the last $\sim20$ days of observation, and a deep transit is clearly visible, with a period of $\sim10$ days. By applying both the PDC and PCA corrections, the spikes are removed, as can be seen from the top-right panel.
In the bottom panel, we focus the attention on the transits. The PDC correction, in the worst cases, could produce several NAN in proximity of the spike in/close to the transits. This effect results in a loss of information that could be crucial if one want to address the properties of the transit. The PCA correction, instead, remove the spikes inside the transit preserving the data point, transit shape and depth.

\begin{figure*}
	\centering
	\includegraphics[scale=0.7]{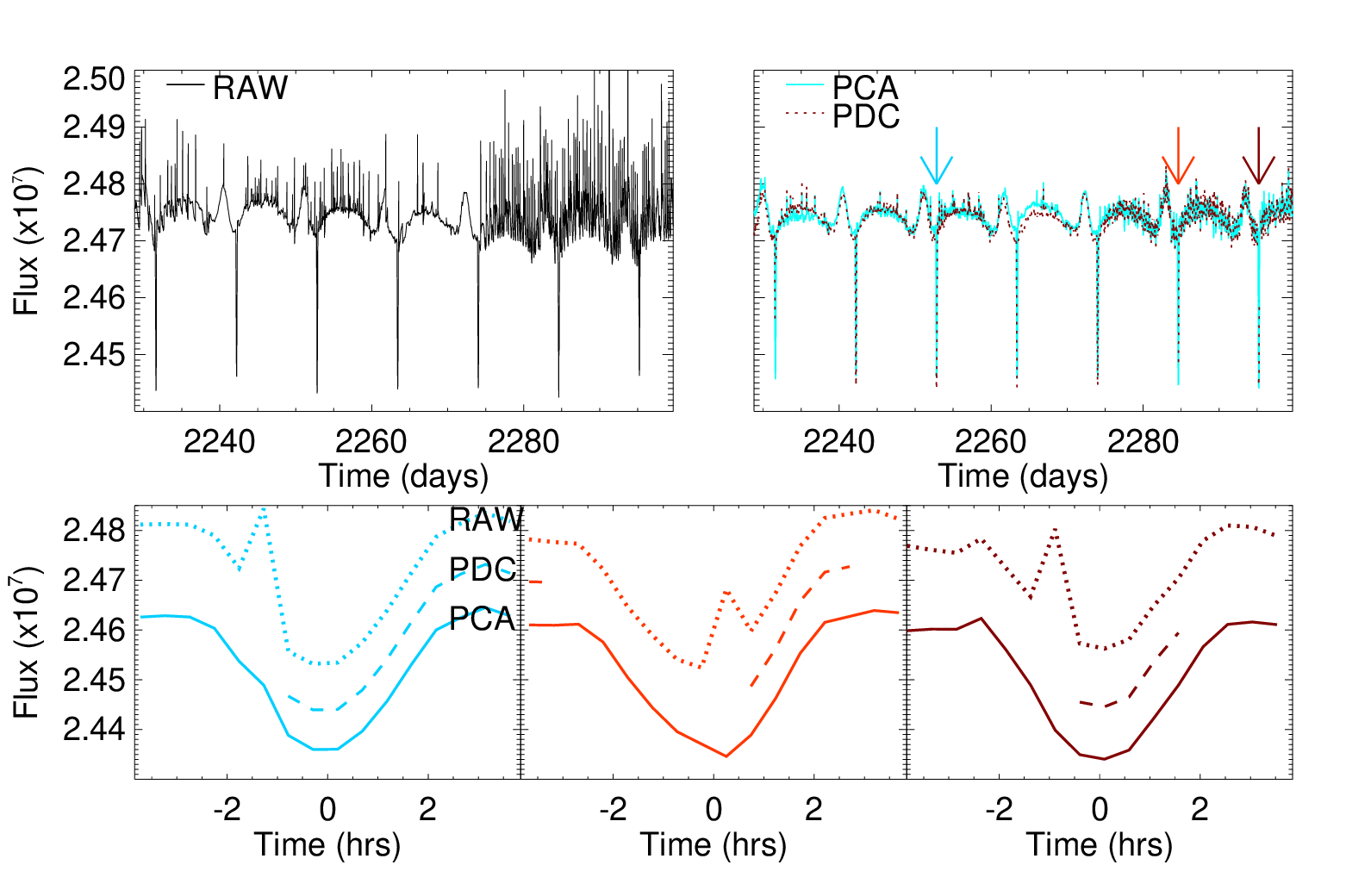}
	
\caption{(Top panel) Light curve in the full observational window ($\sim70$ days), in the case of the original data (left - black line), the PDC corrected data (right - dashed red line) and the PCA corrected one (right - blue line), in which the coloured arrows mark the position of the {three} transits shown in the bottom panel. (Bottom panel) Light curves in a short temporal window ($8$ hrs) around the transits, at the time marked by the arrows in the top panel, and colour coded based on the transit. Each plot shows the original (dotted line), the PDC (dashed line) and the PCA (continuous line) corrected data. For clearness, the original data has been shifted upward by $10^5$ while the PCA corrected one is shifted downward by the same quantity.}
\label{fig:binecl}
\end{figure*}

\subsection{A planet transit around HD3167 (HIP2736, EPIC 220383386)}

Among all the processed light curves, $26$ stars have confirmed transiting planets by June 2019. As an example, we show in Fig.~\ref{fig:plantrans} the light curve of the star HD3167 (HIP2736, EPIC 220383386, Campaign 8). This star has three confirmed orbiting planets, two of them has been revealed by the transit method \cite[HD3167b, HD3167c][]{Vanetal16} while the other by the radial velocity method \cite[HD3167d,][]{Chrisetal17}.

 In top panel we present the light curve in the full observation window for the original data (left) and for the PDC and PCA corrected data (right). As in the case presented in Section~\ref{sec:binecl}, strong spikes overrun the light curve, but the PCA correction is able to remove them. However, we have noticed that, occasionally, the PCA correction {is not able to remove completely some of the strongest spikes, resulting in the presence of artefacts in the processed light curve. This can be due to our choice of the cut-off component that is conservative, however, since they alter the light curve on a very small temporal scale, their effect is not crucial. This can be seen in the bottom panel, in which we show three transits of the planet HD3167b, and in particular in the bottom-left panel as a small triangle-shaped feature,} close to the transit. A cleaning procedure, not implemented in this work, would remove these artefacts. In the other two transits, instead, the spikes are effectively removed.

We have noticed that the PDC procedure could insert an offset in the data in proximity of strong spikes (ex. bottom panel left or middle), but this is not the case for the PCA procedure.

\begin{figure*}
	\centering
	\includegraphics[scale=0.7]{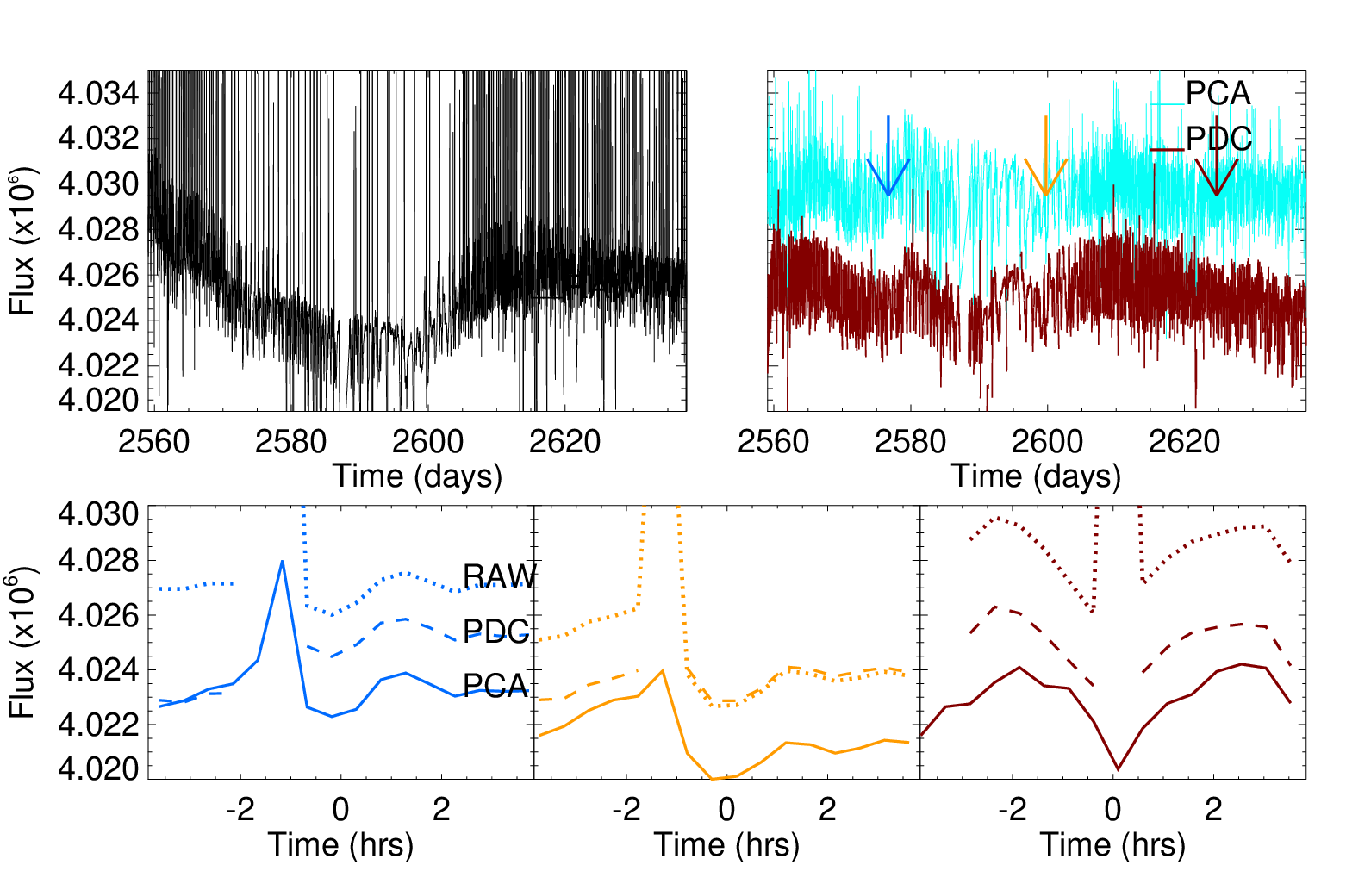}
\caption{ {Same as for Fig.~\ref{fig:binecl} but for the star (HIP2736, EPIC 220383386). In this case the shift applied to the original data and the PCA corrected data in the bottom panels is $2.5\times10^3$ while in the top-right panel the PCA corrected light curve is shifted upward by $4e3$.}}
\label{fig:plantrans}
\end{figure*}

\subsection{Stable sources}
\label{sec:stbsour}
As a further check on the robustness of the method, we have extended the analysis to data set B (stars with a magnitude between $10$ and $11$), for all the campaign described in Sec~\ref{sec:data}, and we have applied the PCA separately to this set of data and to the set obtained by combining it with the data set A previously analysed (magnitude brighter than $10$). 

We are checking on the ability of the method in preserving the intrinsic properties of a star. Here, we consider the subset of stable sources {on the K2 campaign timescale}, therefore, we look for stars that, for at least $95\%$ of the observing time, have variability lower than $10^{-4}$ and $2\times10^{-4}$, relative to their median. The list of stable sources is presented in Tab.~\ref{tab:comp}.

If we look for stable sources at $10^{-4}$, we find $7$ sources in data set A and $1$ in dataset B. If we combine the two data set the number increases to ${12}$. If we move to $2\times10^{-4}$, we find $108$ and $249$ sources, respectively, for the dataset A and B, with the number of stable sources that increases to ${413}$ in the joint dataset. We have checked that stable sources in A and B, when analysed separately, are present in dataset A+B, with only few exceptions, therefore, {since the number of stable sources increases in the joint dataset,} the PCA benefits from larger dataset.

{We have checked if our dataset contains previously selected photometric standards. We have looked for them in different catalogues of photometric standards \citep{Stet2000,LanUm2007,Land2007,Land2009,Land2013,ClemLan2013,BohLan2015,
ClemLan2016} through a cross-match with the GAIA catalogue. We have found only three sources EPIC 201777342 in Campaign 1 (GAIA ID 2537344740061139968), 212688775 in Campaign 6/17 (GAIA ID 3630256618010788864) and 220226402 in Campaign 8 (GAIA ID 2537344740061139968). None of the photometric standards fit out stability thresholds, i.e. variation lower than $10^{-4}$ or $2\times10^{-4}$ with respect to the median for more then the $95$\% of the observational time, but two of them (EPIC 212688775, 220226402) become stable if we set the threshold to $6\times10^{-4}$ while the other is stable at $2\times10^{-2}$. This is compatible with the stability threshold of observations from ground that require maximum magnitude variations in the range $10^{-2}-10^{-3}$ \citep[ex.][]{Stet2000}.}

\begin{table}
\caption{Stable sources grouped by data set, campaign and stability threshold ($10^{-4}$ and $2\times10^{-4}$, {i.e. for more than $95$\% of the time they have a variation relative to their median lower than this threshold}).}
  \label{tab:comp}

\begin{tabular}{ llll } 
\hline\noalign{\smallskip}
   Campaign      & A & B  & A+B   \\ 
\hline     
   {(variation relative to the median)}    &  ($10^{-4}$ / $2\times10^{-4}$) &  ($10^{-4}$ / $2\times10^{-4}$) &  ($10^{-4}$ / $2\times10^{-4}$)\\      
\noalign{\smallskip}\hline\noalign{\smallskip}

      C01     &  $0/0$  &  $0/1$  &  $0/3$   \\ 
      C02     &  $0/0$  &  $0/0$  &  $0/0$   \\ 
      C03     &  $0/4$  &  $0/5$   &  $0/9$  \\ 
      C04     &  $0/3$  &   $0/6$  & $0/8$ \\ 
      C05     &  $2/11$  &  $0/14$    & $2/25$  \\ 
      C06     &  $0/1$  &  $0/12$  & $0/10$   \\ 
      C07     &  $0/2$  &  $0/5$  & $0/10$   \\ 
      C08     &  $0/4$  &  $0/7$  &  $0/10$  \\ 
      C10b     &  $0/0$  &   $0/6$  & $0/4$  \\ 
      C11a     &   $0/16$ &   $0/43$ & $1/69$ \\ 
      C11b     &   $0/10$ &   $0/12$ & $0/22$ \\ 
      C12     &  $0/1$  &   $0/2$    & $0/3$ \\ 
      C13     &  $0/4$  &   $0/7$ &  ${0/11}$   \\ 
      C14     &  $0/6$  &   $0/19$  & $0/29$   \\ 
      C15     &  $0/3$  &   $0/13$  & $0/18$  \\ 
      C16     &  $0/6$  &   $0/3$ & $0/12$   \\ 
      C17     &  $0/3$  &   $0/11$   & $0/17$ \\ 
      C18     &  $2/16$  &  $1/21$ &  ${2/34}$    \\ 
      C19     &  $0/18$  &  $0/62$  & $5/119$  \\ 
\noalign{\smallskip}\hline\noalign{\smallskip}
      Total   &  $7 / 108 $  &  $1/249$ & ${10/413}$  \\ 
\noalign{\smallskip}\hline
\end{tabular}

\end{table}

\section{Discussion and Conclusions}
\label{sec:disc}

In this work, we have developed a procedure that removes systematic effects from a set of light curves on the entire observation time, here applied to K2 data. The procedure makes use of the Principal Component Analysis to identify systematics, and it strongly relies on the individuation of the cut-off component that separates systematics (supposed to explain  most of the data variance) from the astrophysical signal. Once the cut-off component is selected, data reconstruction is a trial process that involves a simple matrix inversion.

We found that the PCA can clean up the light curves from systematics (ex. spikes, offsets) while {it} is able to preserve their intrinsic properties (ex. transit depth and stability/variability). However, the original data treatment (ex. spike treatment) is crucial and it can alter (at very small temporal scales) the results, therefore, our method could benefit from a cleaning procedure at the begin{ning} of the pipeline {(e.g. a $\sigma$-clipping procedure).}

 Another delicate step of the procedure is the selection of the cut-off component. Here, we retrieve it from a fit as three times the cut-off ($\tau$) of the exponential part of the fitting function, for all the campaigns.
 This should be optimised by applying a fine tuning procedure on the selection that can better characterize each campaign. 

The strength of the method rely on the fact we do not need to know any information more than data itself to retrieve the systematics, therefore, it is quite simple to apply to other data from completely different instruments.

One of the first products of the analysis will be the identification of stable sources that can be used as photometric calibrators along the ecliptic plane that can be used for projects as ARIEL that require very stable photometric calibrators.

\begin{acknowledgements}
We acknowledge the support of the ARIEL ASI-INAF agreement n. 2018-22-HH.0
\end{acknowledgements}

\bibliographystyle{spbasic}
\bibliography{pca_expastro_refv1.1_arxiv}

\end{document}